\documentclass[iop]{emulateapj}

\shortauthors{Lapi \& Cavaliere}
\shorttitle{Self-Similar Relaxation of DM Halos}
\slugcomment{Accepted by ApJ}

\begin{document}

\title{Self-Similar Dynamical Relaxation of Dark Matter Halos in an Expanding Universe}

\author{A. Lapi\altaffilmark{1,2} and A. Cavaliere\altaffilmark{1}}
\altaffiltext{1}{Dip. Fisica, Univ. `Tor Vergata', Via Ricerca Scientifica 1,
00133 Roma, Italy.} \altaffiltext{2}{SISSA, Via Bonomea 265, 34136 Trieste,
Italy.}

\begin{abstract}
We investigate the structure of cold dark matter halos using advanced models
of spherical collapse and accretion in an expanding Universe. These base on
solving time-dependent equations for the moments of the phase-space
distribution function in the fluid approximation; our approach includes
non-radial random motions, and most importantly, an advanced treatment of
both dynamical relaxation effects that takes place in the infalling matter:
phase-mixing associated to shell crossing, and collective collisions related
to physical clumpiness. We find self-similar solutions for the
spherically-averaged profiles of mass density $\rho(r)$, pseudo phase-space
density $Q(r)$ and anisotropy parameter $\beta(r)$. These profiles agree with
the outcomes of state-of-the-art $N-$body simulations in the radial range
currently probed by the latter; at smaller radii, we provide specific
predictions. In the perspective provided by our self-similar solutions we
link the halo structure to its two-stage growth history, and propose the
following picture. During the early fast collapse of the inner region
dominated by a few merging clumps, efficient dynamical relaxation plays a key
role in producing a closely universal mass density and pseudo phase-space
density profiles; in particular, these are found to depend only weakly on the
detailed shape of the initial perturbation and the related collapse times.
The subsequent inside-out growth of the outer regions feeds on the slow
accretion of many small clumps and diffuse matter; thus the outskirts are
only mildly affected by dynamical relaxation but are more sensitive to
asymmetries and cosmological variance.
\end{abstract}


\section{Introduction}

\setcounter{footnote}{0}

The cosmogonical paradigm envisages galaxies and galaxy systems to form and
shine when baryons accrete, settle and condense under the gravitational pull
provided by the dark matter (DM), which is cold and collisionless at the
binary level. Understanding the formation history and the detailed structure
of the DM gravitational wells is needed to provide a firm foundation for an
accurate theory of galaxy formation and evolution.

The story starts with DM primordial density perturbations of the cosmic
density field; these at first grow by gravitational instability, and as the
local gravity prevails are enforced to collapse and virialize into
equilibrium `halos'. The resulting halo growth is hierarchical in mass and
sequential in time, with small clumps forming first and then stochastically
merging together into larger and more massive objects.

However, describing and understanding the details of the halo formation
process has proven to be a complex and tricky task, that still challenges the
intense efforts spent by the astrophysical community over the last thirty
years. Major empirical improvements occurred with the advent of the
technology allowing to run intensive $N$-body simulations of many million to
a few billion particles on powerful supercomputers (for a review, see
Springel et al. 2005). A number of hot \emph{issues} are recalled next.

First, DM halos form in \emph{two} stages (Zhao et al. 2003; Diemand et al.
2007; Stadel et al. 2009; Fakhouri, Ma \& Boylan-Kolchin 2010; Genel et al.
2010; Wang et al. 2011): an early fast collapse of the halo bulk including a
few major merger events, which reshuffle the gravitational potential and
cause the DM to undergo (incomplete) dynamical relaxation; a late slow growth
of the halo outskirts in the form of many minor mergers and diffuse
accretion, which little affect the inner potential well, but contribute most
(up to $80\%$) of the final mass.

Second, DM halos in equilibrium show an approximately \emph{universal}
spherically-averaged mass distribution. Originally, this was described with
the simple formula proposed by NFW (Navarro, Frenk \& White 1997), where the
logarithmic density slope $\gamma\equiv -{\rm d}\log\rho/{\rm d}\log r$ is
given by $\gamma = (1+3\,\hat r)/(1+\hat r)$ in terms of the radius $\hat
r\equiv r/r_{-2}$ normalized to the position $r_{-2}$ where $\gamma=2$. This
implies that the density profile $\rho(r)\propto \hat r^{-1}\,(1+\hat
r)^{-2}$ features an inner powerlaw behavior $r^{-1}$, steepens outwards to
$r^{-2}$ in the halo middle, and then goes into an asymptotic shape $r^{-3}$
in the outskirts. Remarkably, the NFW formula was found to be approximately
scale-invariant, i.e., to fit accurately the simulation outcomes for halos of
different mass scales. On the other hand, a weak scale dependence is
introduced by the concentration parameter $c\equiv R_{200}/r_{-2}$, that is
the ratio between $r_{-2}$ and the radius $R_{200}$ where the overdensity
relative to the background amounts to $200$; despite the name, it actually
constitutes a measure of the halo outer \emph{extension}. Simulations show
that $c\approx 3.5$ applies at the end of the fast collapse (with minor mass
and redshift dependencies, see Prada et al. 2011), and increases as $c\propto
H(z)\propto (1+z)^{-1}$ afterwards. At the present time $z\approx 0$, a
relation $c\propto M^{-0.13}$ applies, with a scatter around $0.2$ dex due to
variance in the growth histories (see Bullock et al. 2001); typically, a
galaxy halo with current mass $M\approx$ a few $10^{12}\, M_\odot$ collapsed
at $z\approx 2$ features a concentration value $c\approx 10$, while the halo
of a galaxy cluster with $M\approx 10^{15}\, M_\odot$ collapsed at $z\approx
0.5$ features a value $c\approx 5$.

In addition, recent simulations (Navarro et al. 2004, 2010) have shown
evidence of small but systematic deviations from the simple NFW expression;
they rather favor a S\'{e}rsic-Einasto (S\'{e}rsic 1963, Einasto 1965; see
also Prugniel \& Simien 1997; Merritt et al. 2006; Lapi \& Cavaliere 2011)
formula $\gamma=\tau+(2-\tau)\,\hat r^\eta$, depending on the two parameters
$\tau$ and $\eta$ that describe the inner behavior and the curvature of the
density profile $\rho(r)\propto r^{-\tau}\, e^{-(2-\tau)\, r^{\eta}/\eta}$
before a final exponential cutoff. Current simulations set the upper limit
$\tau\la 0.9$ on the inner asymptotic powerlaw but still lack enough
resolution to pinpoint the true value. On the other hand, if a pure Einasto
(1965) model with $\tau=0$ is adopted, different values of $\eta\approx
0.1-0.2$ are required to provide precise fits to different simulated halos.
This indicates that scale-invariance in the mass distribution is actually
broken, and/or that a halo's development is appreciably affected by variance
related to detailed merging histories or environmental conditions.

Third, DM halos show a definite spherically-averaged profile of anisotropy.
This is described on using the standard Binney (1978) parameter
$\beta(r)\equiv 1-\sigma_\theta^2(r)/\sigma_r^2(r)$ in terms of the radial
and tangential velocity dispersions $\sigma_{r\,,\,\theta}^2(r)$. Halos are
found to be quasi-isotropic at the center with $\beta\approx 0$ within the
resolution limits, and to become radially anisotropic outwards. Specifically,
$\beta\approx 0.25$ applies at $r\approx r_{-2}$ and $\beta\approx 0.5$ at
$r\approx$ a few $r_{-2}$; then $\beta$ decreases outwards, though with wide
oscillations (e.g., Navarro et al. 2010; Ludlow et al. 2010). Remarkably,
within $r_{-2}$ the anisotropy parameter and the slope of the density profile
appear to be correlated (see Huss, Jain \& Steinmetz 1999; Hansen \& Moore
2006) through a $\beta-\gamma$ relation of approximate form $\beta(r)\approx
-0.15+0.2\, \gamma(r)$.

Finally, DM halos show a powerlaw spherically-averaged profile of the
quantity $Q(r)\equiv \rho/\sigma^3$. This has the dimensions of a phase-space
density, but it is not a true measure neither of it nor of its coarse-grained
version (see discussion by Ascasibar \& Binney 2005; Sharma \& Steinmetz
2006); thus it is often referred to as a pseudo (or a proxy of the)
phase-space density, although it is still debated whether the radial or the
total velocity dispersion should enter its definition (see discussion by
Schmidt, Hansen \& Macci\'{o} 2008). Remarkably, although the density
$\rho(r)$ and velocity dispersion $\sigma^2(r)$ have articulated runs, $Q(r)$
follows a simple powerlaw $Q(r)\propto r^{-\chi}$ over three order of
magnitude in radius, with the same exponent $\chi\approx 1.9$ applying for
all halos (Taylor \& Navarro 2001; Hoffman et al. 2007; Ascasibar \&
Gottl\"{o}ber 2008; Vass et al. 2009; Navarro et al. 2010; Ludlow et al.
2010). On the other hand, the recent simulations highlight that such a
powerlaw behavior holds within $r_{-2}$, but that $Q(r)$ possibly steepens in
the inner regions, while it flattens appreciably in the outskirts. The origin
of the powerlaw behavior is presently unknown, but numerical experiments with
different perturbation spectra indicate that it is not related to initial
conditions or hierarchical merging (see Wang \& White 2009).

Unfortunately, these remarkable findings still lack a firm theoretical
background, while their reliability is restricted in the inner and outer
regions of the halos by limited resolution and small particle statistics,
respectively. These are good reasons to complement the numerical approach
with \emph{analytic} models.

In the analytic vein, a simple approach is provided by the self-gravitating,
static equilibria of DM based on the Jeans equation (Taylor \& Navarro 2001;
Austin et al. 2005; Dehnen \& McLaughlin 2005; Lapi \& Cavaliere 2009a,
2009b, 2011)
\begin{equation}
{1\over \rho}\,{\mathrm{d}\over \mathrm{d}
r}\,(\rho\sigma_r^2)+2\,\beta\,{\sigma_r^2\over r}+{G M(<r)\over r^2}=0~.
\end{equation}
Two ingredients are needed to solve it for $\rho(r)$: the profile of the
anisotropy parameter $\beta(r)$, and that of the phase-space density $Q(r)$
linking $\rho(r)$ and $\sigma_r^2(r)$ in the spirit of an `equation of
state'. Both runs can be predicted from simple scaling arguments, and refined
by comparison with numerical simulations (see above), to the effect that
$\beta(r)$ takes the form of the linear $\beta-\gamma$ relation, while
$Q(r)\propto r^{-1.9}$ applies. Interestingly, the resulting density profiles
are well described by the S\'{e}rsic-Einasto formula (Lapi \& Cavaliere
2011), and turn out to be very close to the simulation results in the halo
middle where the Jeans solutions are reliable; conversely, Ludlow et al.
(2011) take up from simulations an Einasto shape for $\rho(r)$ and solve
Jeans to closely recover the powerlaw run of $Q(r)$. However, such analytic
approaches are limited being based on a static equation; they provide final
equilibrium pictures of halos, but little can tell on how DM particles
progressively collapse or accrete, and then relax. As such, they sidestep
origins and building up of the profiles $Q(r)$ and $\beta(r)$, and are
expected to fail close to the halo outskirts, in the region exposed to infall
where equilibrium is not yet achieved.

To go beyond these limitations requires formulating \emph{physical} models of
infall and accretion in an expanding universe; to that purpose we resort to
an advanced fluid-like description of the DM dynamics including random
non-radial motions and relaxation effects, but still amenable to a
self-similar treatment. The latter provides a tractable, analytic way of
investigating time-dependent problems in complex physical systems; it is
possible whenever the system dynamics can be characterized, besides spacetime
variables, by a few parameters with independent dimensions. Although
providing only a particular analytic solution to the physical problem, it
often accurately yields the long-time behaviors and offers a useful guide for
understanding the generic features of the system (see discussions by Sedov
1959; Zel'dovich \& Raizer 1967). In the present context, a self-similar
description for the collapse and condensation of a DM halo is allowed by
choosing a scale-invariant shape of the initial mass perturbation, and
adopting an Einstein-de-Sitter cosmological framework.

Self-similar solutions for the purely radial infall of collisionless matter
have been pioneered by Gunn \& Gott (1972), then explicitly derived by
Fillmore \& Goldreich (1984) and Bertschinger (1985) on adopting a Lagrangian
description for the orbits of individual particles, and by Teyssier et al.
(1997) and Subramanian (2000) on adopting an equivalent fluid approximation
for the ensemble of particles. Such self-similar treatments with purely
radial infall concur in yielding very steep inner density profiles with
$\gamma\sim 2$, at variance with the simulation outcomes.

White \& Zaritsky (1991), Ryden (1993), Sikivie et al. (1997), Subramanian
(2000), Nusser (2001), Zukin \& Bertschinger (2010a, 2010b), Vogelsberger et
al. (2011), and Lithwick \& Dalal (2011) took steps toward fixing the problem
by inclusion of non-radial motions originated by tidal torques, either in
spherical or triaxial collapses; when angular momentum is present, the system
develops a tangential velocity dispersion that causes a flattening of the
inner density profile (see also Eq.~1). However, the quantitative effect is
sensitive to the specific amount of angular momentum endowed at, or acquired
during the infall, so the solutions still fail to explain the approximately
universal shapes found in simulations (see discussion by Lu et al. 2006).

The \emph{view} we submit here envisages that all such shortcomings go back
to a key ingredient missing (or at least, inadequately treated) in the
classic self-similar models, i.e., dynamical relaxation. This involves two
different mechanisms leading to broadly similar macroscopic outcomes:
phase-mixing, a localized process related to spreading of neighboring
particle orbits over phase-space; and violent relaxation, a volume process
driven by irregular fluctuations of the gravitational potential. Both these
effects have long been recognized to play a crucial role in the approach of a
gravitational system to its equilibrium configuration, with violent
relaxation especially relevant when the initial condition of the collapse is
clumpy (e.g., van Albada 1982; Ma \& Bertschinger 2004; Lu et al. 2006). This
is certainly the case in the standard cosmological framework, where DM
objects form from accretion punctuated by mergers.

For example, Lynden-Bell (1967) and Nakamura (2000) based on a statistical
mechanical approach to show that an equilibrium state may be achieved when
violent, full relaxation erases the memory of the initial conditions. Besides
delicate issues concerning the mathematical consistency of these theories
(see discussion by Arad \& Lynden-Bell 2005), a number of numerical
experiments have demonstrated that the relaxation is always incomplete (see
van Albada 1982; Henriksen \& Widrow 1999; Trenti, Bertin \& van Albada 2005,
and references therein), with a significant correlation retained between the
final and initial state of a particle. In fact, this may explain why the
detailed density profiles of virialized DM halos ultimately do depend on the
formation history and on cosmogonical conditions.

In this paper we propose an advanced treatment of both dynamical relaxation
effects within self-similar solutions for the spherical collapse of DM halos
in a fluid-like approach. We will show that our description resolves the
discrepancy between the outcomes of self-similar models and numerical
simulations, in that the solutions of the former feature profiles of mass
density, pseudo phase-space density and anisotropy consistent with the
latter.

The plan of the paper is straightforward. In \S~2 we derive the evolution
equations for the moments of the DM distribution function in the fluid
approximation, including the terms responsible for dynamical relaxation (see
Appendix). In \S~3 we reduce these equations to a self-similar form. In \S~4
we derive conservation laws, investigate the inner asymptotic behaviors of
the solutions, present the full solutions, and compare them with the
simulation outcomes. Finally, in \S~5 we discuss our approach and summarize
our main conclusions.

\section{DM dynamics in the fluid approximation}

We consider a spherically symmetric, nonrotating DM halo, and describe
particle positions and velocities in terms of the standard polar coordinates
$r$, $\theta$, $\phi$, and of the corresponding velocity components $v_r$,
$v_\theta$, $v_\phi$. In the phase-space the particle distribution function
$f(r,v_r,v_\theta,v_\phi)$ evolves as dictated by the classic Boltzmann
equation\footnote{For the sake of simplicity, in spherical symmetry and in
the absence of net macroscopic rotation of the halo, one can neglect any
explicit dependence of $f$ on the angular variables $\theta$ and $\phi$ (see
Larson 1969, 1970; Subramanian 2000). In fact, the related terms
$(\partial_\theta f)/r$, $(\partial_\phi f)/r\,\sin\theta$ to be added on the
l.h.s. of Eq.~(2) would introduce in the moment equations (Eqs.~5 to be
derived next) averages like $<v_{\theta\,,\,\phi}>$ or dispersions like
$\sigma^2_{r\theta\,,\,r\phi\,,\,\theta\phi}$ that must vanish anyway in a
macroscopically spherical, nonrotating halo.}
\begin{equation}
\partial_t\,f+v_r\,\partial_{r}\, f+\dot v_r\,\partial_{v_r}\, f+\dot
v_\theta\,\partial_{v_\theta}\,
f+\dot v_\phi\,\partial_{v_\phi}\, f=(\partial_t\, f)_\star~.
\end{equation}
Here $(\partial_t\, f)_\star$ is a term that describes collective
`collisions' contributing to dynamical relaxation, and will be specified in
\S~2.1 and in the Appendix. The single-particle velocity components change
according to the equations of motion
\begin{eqnarray}
\nonumber \dot v_r &=&-{G\, M(<r)\over r^2}+{v_\theta^2+v_\phi^2\over r}~,\\
\nonumber \\
\dot v_\theta &=& -{v_r\,v_\theta\over r}+{v_\phi^2\over r\, \tan{\theta}}~,\\
\nonumber \\
\nonumber \dot v_\phi &=& -{v_r\,v_\phi\over r}-{v_\theta\,v_\phi\over r\,
\tan{\theta}}~,
\end{eqnarray}
where $M(<r)$ is the DM mass within a radius $r$.

Next we derive the evolution equations for moments of the velocity
distribution, up to the second order inclusive. In the following the `mean'
value $<X>$ of a generic quantity $X$ is defined by
\begin{equation}
<X>\equiv {1\over \rho}\, \int{\rm d}^3\,v\,f\, X~,
\end{equation}
where the spatial density is $\rho\equiv \int{\rm d}^3\,v\,f$. In the
tangential directions, the mean components of the velocity
$<v_{\theta\,,\,\phi}>=0$ must vanish (as all other moments of odd order)
while the corresponding dispersions $\sigma_\theta^2=\sigma_\phi^2$ must be
equal (in fact, the value of any moment is unaffected by interchanging
$v_\theta$ and $v_\phi$). In the radial direction, instead, one has to
consider the mean velocity $u\equiv <v_r>$ and the dispersion
$\sigma_r^2\equiv <(v_r-u)^2>=<v_r^2>-u^2$.

To obtain a closed set of moment equations we need to relate the third moment
with the lowest-order ones. The standard and simplest assumption is to
require zero radial skewness $<(v_r-u)^3>=0$, which yields
$<v_r^3>=u\,(u^2+3\,\sigma_r^2)$; in the literature this is often referred to
as `\emph{fluid} approximation' (see Teyssier et al. 1997; Chuzhoy \& Nusser
2000; Subramanian 2000; also Dehnen \& Read 2011 and references therein).
Radial skewness, on the other hand, would imply an outward flow of energy
through the system, or in other words a tendency for the most energetic
particles to move preferentially outwards. Skewness is often taken into
account in studies of star clusters' outskirts to evaluate the secular
evolution of the system due to the escape of energetic stars; on the other
hand, two-body encounters tend to erase skewness at the center over a
timescale slightly longer than the local relaxation time, see Lynden-Bell \&
Wood (1968) and Larson (1970). These analyses suggest skewness to be smaller
than second-order moments by one order of magnitude or more throughout the
system. In the context of DM halos accreting from a cosmological environment,
it has been shown (see references at the beginning of the paragraph) that
skewness is negligible in the inner region where large amount of phase-mixing
has occurred, and it may become relevant only in a narrow region around the
virial boundary (or better, the outermost caustics), which in the fluid
approximation is considered as a sharp discontinuity and dealt with as
discussed in \S~3.1.

Multiplying Eq.~(2) by $1$, $v_r$, $(v_r-u)^2$, $v_\theta^2$ and integrating
over the velocity distribution yields the system of coupled partial
differential equations
\begin{eqnarray}
\nonumber &&(\partial_t+u\,\partial_r)\,\rho+{\rho\over r^2}\,\partial_r\,(r^2\,u)=0~,\\
\nonumber \\
\nonumber &&(\partial_t+u\,\partial_r)\,u+{1\over \rho}\,\partial_r\,(\rho\,\sigma_r^2)+{2\over
r}\,(\sigma_r^2-\sigma_\theta^2)+{G\, M(<r)\over r^2}=0~,\\
\nonumber \\
&&(\partial_t+u\,\partial_r)\,\sigma_r^2+2\,\sigma_r^2\,\partial_r\,u=(\partial_t\,\sigma_r^2)_\star~,\\
\nonumber\\
\nonumber &&(\partial_t+u\,\partial_r)\,\sigma_\theta^2+{2\over
r}\,\sigma_\theta^2\,u=(\partial_t\,\sigma_\theta^2)_\star~,\\
\nonumber\\
\nonumber &&\partial_r M=4\pi\,r^2\,\rho~.
\end{eqnarray}
These are commonly referred to as the continuity, momentum, energy, angular
momentum, and mass equations; note that the second constitutes the extension
of Eq.~(1) to time-dependent conditions.

Remarkably, the first and the third Eqs.~(5) can be combined into the form
\begin{equation}
(\partial_t+u\,\partial_r)\,\log{\sigma_r^2\over
\rho^2\,r^4}={(\partial_t\,\sigma_r^2)_\star\over \sigma_r^2}~.
\end{equation}
This describes an entropy flow for a one-dimensional fluid with effective
density $\tilde\rho\equiv\rho\,r^2$, effective pressure $\tilde
p\equiv\tilde\rho\,\sigma_r^2$, and effective specific entropy $\tilde
s\equiv \log \tilde p/\tilde \rho^3$, in terms of a microscopic adiabatic
index equal to $3$.

\subsection{The `collision' terms}

We now turn to discuss the terms
$(\partial_t\,\sigma^2_{r\,,\,\theta})_\star$ on the r.h.s. of the third and
fourth Eqs.~(5), that describe collective `collisions' in the DM fluid;
hereafter these will be simply referred to as collision terms. Physically,
they arise because clumpiness in the infalling matter induces fluctuations of
the gravitational potential that \emph{collectively} affect the dynamics of
DM particles. In fact, a particle effectively sees a stochastically
fluctuating potential, and (consistent with the fluctuation-dissipation
theorem) suffers of a dissipative drag that produces an effective `cooling'
(see Chandrasekhar 1943; Kandrup 1980; Antonuccio-Delogu \& Atrio-Barandela
1992; Del Popolo \& Gambera 1997; Ma \& Bertschinger 2004; for a review see
Binney \& Tremaine 2008). On the other hand, as pointed out by Bekenstein \&
Maoz (1992) and Ma \& Boylan-Kolchin (2004), effective `heating' of a
particle by the fluctuations can also constitute a relevant process.

In the present context, we consider a halo of mass $M$ including clumps of
typical mass $m$. In Appendix A we work out the form of the collision terms
in the Fokker-Planck approximation, under the assumption of a quasi-isotropic
distribution function; the results are as follows
\begin{eqnarray}
\nonumber && (\partial_t\, \sigma_r^2)_\star=-{64\,\sqrt{3\,\pi}\over
15}\,G^2\,m\,\rho\,\log\Lambda\,{\sigma_r^2-\sigma_\theta^2\over \sigma_{\rm tot}^3}~,\\
\\
\nonumber && (\partial_t\, \sigma_\theta^2)_\star=+{32\,\sqrt{3\,\pi}\over
15}\,G^2\,m\,\rho\,\log\Lambda\,{\sigma_r^2-\sigma_\theta^2\over \sigma_{\rm
tot}^3}~,
\end{eqnarray}
with $\log \Lambda\equiv \log M/m$ the standard `Coulomb logarithm' (see
Appendix A.2 for details). Note that the ratio of the average relaxation
timescale $\sim 10^{-1}\,\sigma^3/G^2\, m\, \rho\, \log(M/m)$ from Eqs.~(7)
to the Hubble time $(3/8\pi\, G\, \rho)^{1/2}$ reads $\sim$ a few $10^{-1}\,
\mathcal{N}/\log \mathcal{N}$ in terms of the effective clump number
$\mathcal{N}\equiv M/m$; this implies that relaxation can be efficient for
$\mathcal{N}\la 10$, i.e., when a limited number of clumps is present (a more
detailed discussion of the relaxation strength is provided in Appendix A.2
and recalled in \S~3).

Note that in Eqs.~(7) the (negative) cooling term dominates in the radial
direction, while the (positive) heating term dominates in the tangential
direction. The collisions induced by clumps sharing the accretion inflow do
not change the total velocity dispersion $\sigma^2_{\rm tot}\equiv
\sigma_{r}^2+2\,\sigma_{\theta}^2$, since $(\partial_t\,\sigma_{\rm
tot}^2)_\star=0$ holds; this means that the overall random component of the
energy is conserved, being just redistributed between the radial and the
tangential degrees of freedom. On the other hand, collisions tend to
\emph{erase} any velocity anisotropy since
$[\partial_t\,(\sigma_{r}^2-\sigma_\theta^2)]_\star\propto
-(\sigma_{r}^2-\sigma_\theta^2)$ holds.

Such collective collisions provide one mechanism contributing to dynamical
relaxation, close to the classic but still incomplete notion of violent
relaxation (see \S~1 and references therein). The other mechanism is related
to phase-mixing, and its implications in the present context will be
discussed in \S~3.1.

\section{Self-similar description}

Self-similar solutions of the above Eqs.~(5) obtain under two assumptions:
(i) an Einstein-de-Sitter cosmological framework, which still provides an
increasingly good approximation to the concordance cosmology for $z\ga 0.5$
where most galactic halos form and evolve; (ii) a spherically-symmetric,
scale-free shape of the initial DM mass perturbation of the form
\begin{equation}
{\delta M\over M}\propto M^{-\epsilon}\propto r^{-3\,\epsilon}~,
\end{equation}
which may be considered as a piecewise approximation to a realistically
bell-shaped cold DM perturbation. Here $\delta M$ represents the mass excess
in a shell of initial comoving radius $r_i\propto M^{1/3}$ enclosing a mass
$M$ of matter at background density. Such a shell will progressively detach
from the Hubble flow, reach a maximum `turnaround' radius $R_{\rm ta}\propto
r_i/(\delta M/M)\propto M^{\epsilon+1/3}$, and collapse back to the standard
virial radius $R_{200}\la R_{\rm ta}/2$.

The virialization occurs when $\delta M/M$ attains the critical threshold
$1.686\, D^{-1}(t)$ in terms of the linear growth factor $D(t)\propto
t^{2/3}$ along the cosmic time $t$ in the Einstein-de-Sitter cosmology. So
the shape parameter $\epsilon$ also governs the mass buildup after
$M(t)\propto D^{1/\epsilon}(t)\propto t^{2/3\,\epsilon}$, or equivalently
sets the collapse time $t_{\rm coll}\equiv M/\dot M=3\, \epsilon\,t/2$ for
the shell surrounding the mass $M$. Values $\epsilon\la 2/3$ correspond to a
stage of fast collapse relative to the Hubble expansion, while values
$\epsilon\ga 2/3$ correspond to a slow mass accretion.

The scaling $M\propto t^{2/3\epsilon}\propto (1+z)^{-1/\epsilon}$ may be
conveniently compared with that of the characteristic clustering mass
$M_c\propto (1+z)^{-6/(n+3)}$ in a scale-free hierarchical cosmology with a
power spectrum $P(k)\propto k^n$ (e.g., Peebles 1980). Perturbations
characterized by a specific value of $\epsilon$ therefore accrete mass at the
same rate as the `typical' mass with $\epsilon=(n+3)/6$. For example, the
fast collapse of the perturbation bulk for a galactic halo is described by
$\epsilon\approx 1/12\div 1/6$ corresponding to $n\approx -2.5\div -2$, while
for the halo of a galaxy cluster values $\epsilon\approx 1/6\div 1/3$ apply
corresponding to $n\approx -2\div -1$. On the other hand, during the
development of the outskirts, the accretion rate slows down and $\epsilon$
can grow considerably larger (see discussion in \S~5).

Once the shape of the initial perturbation has been specified, there are no
additional physical scales in the problem, and the time evolution of the
system must approach \emph{self-similarity} after a short initial transient.
This implies that a single solution describes the structure and time behavior
of the system, when expressed in properly scaled variables. Since the current
turnaround radius is easily seen to grow as
\begin{equation}
R_{\rm ta}(t)\propto t^{\xi}~,~~~~{\rm with }~~~~\xi\equiv {2\over
9\epsilon}\,(1+3\epsilon)~,
\end{equation}
it is convenient to introduce the self-similar variable $\lambda\equiv
r/R_{\rm ta}(t)$, and define the adimensional radial velocity, density, mass,
and velocity dispersions through the relations
\begin{eqnarray}
\nonumber &&u(r,t)={R_{\rm ta}\over t}\,\mathcal{U}(\lambda)~,\\
\nonumber \\
\nonumber &&\rho(r,t)={1\over 6\pi\,G\,t^2}\,\mathcal{D}(\lambda)~,\\
\\
\nonumber &&M(<r,t)={2\,R_{\rm ta}^3\over 9\,G\,t^2}\,\mathcal{M}(\lambda)~,\\
\nonumber\\
\nonumber &&\sigma^2_{r\,,\,\theta}(r,t)= {R_{\rm ta}^2\over
t^2}\,\Sigma^2_{r\,,\,\theta}(\lambda)~.
\end{eqnarray}

Thus we can transform Eqs.~(5) into the system of ordinary differential
equations
\begin{eqnarray}
\nonumber &&
(\mathcal{U}-\lambda\,\xi)\,\mathcal{D}'+\left({2\,\mathcal{U}\over
\lambda}+\mathcal{U}'-2\right)\,\mathcal{D}=0~,\\
\nonumber \\
\nonumber
&&(\mathcal{U}-\lambda\,\xi)\,\mathcal{U}'+(\xi-1)\,\mathcal{U}+{1\over
\mathcal{D}}\,(\mathcal{D}\,\Sigma_r^2)'+\\
&&  ~~~~~~~~~~~~~~~+{2\over \lambda}\,(\Sigma_r^2-\Sigma_\theta^2)+{2\over 9}\,{\mathcal{M}\over
\lambda^2}=0~,\\
\nonumber\\
\nonumber
&&(\mathcal{U}-\lambda\,\xi)\,(\Sigma_r^2)'+2\,\left(\xi-1+\mathcal{U}'\right)\,\Sigma_r^2=(\Sigma_r^2)_\star'~,\\
\nonumber\\
\nonumber
&&(\mathcal{U}-\lambda\,\xi)\,(\Sigma_\theta^2)'+2\,\left(\xi-1+{\mathcal{U}\over\lambda}\right)\,\Sigma_\theta^2=(\Sigma_\theta^2)_\star'~,\\
\nonumber\\
\nonumber &&\mathcal{M}'=3\,\lambda^2\,\mathcal{D}~,
\end{eqnarray}
where prime denotes differentiation with respect to $\lambda$.

Moreover, with a constant value for the effective number $\mathcal{N}\equiv
M/m$ of clumps in the infalling matter, the collision terms are seen to scale
as $1/t$, and hence they do not break the self-similarity. Using the
variables defined in Eqs.~(10) we can write them in the form
\begin{eqnarray}
\nonumber && (\Sigma_r^2)_\star'=-\kappa_\star\, {\mathcal{D}\,
\mathcal{M}\over \Sigma_{\rm tot}^3}\,
(\Sigma_r^2-\Sigma_\theta^2)~,\\
\\
\nonumber && (\Sigma_\theta^2)_\star'=+{\kappa_\star\over 2}\, {\mathcal{D}\,
\mathcal{M}\over \Sigma_{\rm tot}^3}\, (\Sigma_r^2-\Sigma_\theta^2)~,
\end{eqnarray}
ready to be inserted in Eqs.~(11). In the above expression
\begin{equation}
\kappa_\star\equiv {64\,\sqrt{3\pi}\over 405}\,{\log \mathcal{N}\over
\mathcal{N}}~
\end{equation}
is the strength parameter of the collective collisions, which depends mainly
on the `clumpiness' $1/\mathcal{N}$ of the infalling matter, see
Appendix A and Fig.~A1 for details.

In solving numerically the fully self-similar Eqs.~(11), we will consider
different values of $\kappa_\star$ guided by the following physical
considerations. During the fast accretion ($\epsilon\la 2/3$) we expect that
a limited number of major clumps $\mathcal{N}\la 10$ rapidly merge to build
up the halo bulk; after Eq.~(13) this implies \emph{efficient} dynamical
relaxation, with strength parameter taking on values $\kappa_\star\approx
0.1$. On the other hand, during the late development of the outskirts we
expect slow accretion ($\epsilon\ga 2/3$) of many small clumps with
$\mathcal{N}\ga 30$; these conditions imply \emph{inefficient} dynamical
relaxation with $\kappa_\star\la 0.01$.

We stress that such values for the effective number of clumps in the
different accretion regimes are consistent with the findings of numerical
simulations (see Wang et al. 2011, their Fig.~7). The latter show that most
of the mass in the inner region (which in turn is about $20\%$ of the total)
is contributed by a number of $5-10$ major mergers with mass ratios
$1:2-1:3$; on the other hand, most of the mass in the outskirts is
contributed by a number $\ga 30$ of minor mergers with mass ratio $\la 1:10$,
and the rest by smooth and diffuse accretion.

\subsection{Boundary conditions}

Since the original Eqs.~(5) describing the DM flow are hyperbolic,
discontinuities are expected to develop. However, given the collisionless
nature of the DM particles (at the binary level), the discontinuities are
constituted not by shocks, but rather by caustics where the bulk infall
energy is partly converted into random motions. The caustics occur very close
to the turning points where the radial velocity of a shell vanishes (e.g.,
Bertschinger 1985); there adjacent mass shells catch up with each other, and
the intervening matter is compressed to a divergent density. Inward of a
caustic multiple `shell crossings' occur, the effective gravitational force
experienced by a particle is (albeit slowly) time dependent, and neighboring
particles' orbits go out of phase, causing the so called `phase-mixing'.
Deeper in radius and/or later in time, more shell crossings have occurred, to
originate a smoother coarse-grained particle distribution in phase-space.

In the treatments based on the Lagrangian viewpoint where the orbits of
single DM particles are followed, phase-mixing primarily takes place in an
outer layer including a few caustics (see Fillmore \& Goldreich 1984,
Bertschinger 1985). In the fluid approximation one renders the layer of such
caustics with a single `\emph{discontinuity}' across which the relevant
Rankine-Hugoniot$-$type jump conditions apply. An extensive literature has
shown the effectiveness of such an approach in closely matching the
single-particle results (see Teyssier et al. 1997; Chuzhoy \& Nusser 2000;
Subramanian 2000). On the other hand, in the volume inward of the
discontinuity the other relaxation mechanism related to collective collisions
can operate (see \S~2.1).

The radius where the caustic discontinuity occurs in the self-similar
description must be a constant fraction $\lambda_c$ of the turnaround radius,
and will be computed below as an eigenvalue (see Bertschinger 1985). In
detail, we proceed as follows. First of all, outside $\lambda_c$ the
evolution is identical to the turnaround and collapse of a pressureless mass
shell; a parametric form of the upstream accretion flow is given by (e.g.,
Peebles 1980)
\begin{eqnarray}
\nonumber && \lambda=\sin^2{\theta\over 2}\,\left({\theta-\sin{\theta}\over
\pi}\right)^{-\xi}~,\\
\nonumber \\
\nonumber &&
\mathcal{U}(\lambda)={\lambda\sin{\theta}\,(\theta-\sin{\theta})\over
(1-\cos{\theta})^2}~,\\
\\
\nonumber&&\mathcal{D}(\lambda)={9\over 2}\,{(\theta-\sin{\theta})^2\over
(1-\cos{\theta})^3\,(1+3\epsilon\chi)}~~~~{\rm
with}~~\chi=1-{3\over 2}\,{\mathcal{U}(\lambda)\over \lambda}~,\\
\nonumber\\
\nonumber &&\mathcal{M}(\lambda)={9\over
2}\,\lambda^3\,{(\theta-\sin{\theta})^2\over (1-\cos{\theta})^3}~.
\end{eqnarray}
Note that at the turnaround position corresponding to $\lambda=1$ the above
Eqs.~(14) give $\mathcal{U}(1)=\Sigma_{r\,,\,\theta}^2(1)=0$,
$\mathcal{M}(1)=(3\pi/4)^2$ and
$\mathcal{D}(1)=\mathcal{M}(1)/(1+3\epsilon)$.

Then, by integrating the self-similar Eqs.~(11) in a region across the
caustic discontinuity, we obtain the \emph{jumps}
\begin{eqnarray}
\nonumber &&\mathcal{U}_+-\xi\,\lambda_c={1\over 2}\,(\mathcal{U}_--\xi\lambda_c) ~,\\
\nonumber \\
&&\mathcal{D}_+=2\,\mathcal{D}_-~,\\
\nonumber\\
\nonumber &&(\Sigma_r^2)_+={1\over 16}\,(\mathcal{U}_--\xi\lambda_c)^2~,
\end{eqnarray}
here the $-$, $+$ signs refer to upstream and downstream values,
respectively. Outward of the discontinuity, we have taken the radial and
tangential velocity dispersion to be null, since only bulk radial motions are
associated to the inflow of a shell until it collapses back to $\lambda_c$
and undergoes shell crossing. At the discontinuity, the flow is compressed
and the bulk radial velocity is converted into radial dispersion, similarly
to what occurs with the usual jump conditions at a shock but for a fluid with
one degree of freedom.

On the other hand, tangential velocity dispersion is also generated by a
nonspherical collapse, but a simple treatment like the present one cannot
tell to what degree this occurs; in fact, no source term for the tangential
dispersion is present in the corresponding moment equation, meaning that it
must be assigned as a boundary condition. For example, Zukin \&
Bertschinger (2010a) took steps toward estimating the tangential dispersion
at the turnaround basing on tidal torque theory; their result overestimates
somewhat (by a factor about $2$) the outcomes of $N-$body simulations, and
ultimately must be tuned in terms of the latter. This issue is beyond the
scope of the present paper, and we just parameterize the tangential
dispersion at the caustic discontinuity in terms of the boundary value
$\beta_+\equiv 1-(\Sigma_\theta^2/\Sigma_r^2)_+$ for the anisotropy
parameter. We shall find that such a value does not affect the \emph{inner}
shapes of the mass and pseudo phase-space density profiles when efficient
dynamical relaxation is at work.

Technically, the location of the caustic discontinuity $\lambda_c$ is
determined as an eigenvalue (e.g., Bertschinger 1985), by imposing the
following inner physical constraints
\begin{equation}
\mathcal{M}(0)=\mathcal{U}(0)=0~,
\end{equation}
i.e., the mass and velocity at the center must vanish. Note that the caustic
constitutes an effective\emph{ boundary} for the halo, which is close if
exterior to the virial radius $R_{200}$.

\section{Self-similar solutions}

We now turn to solving the self-similar Eqs.~(11). Before handling the
problem numerically, useful insights into the behavior of the solutions is
found from the analytic work that follows.

\subsection{Conservation laws}

First of all, we find that integrals of motion are associated to the
self-similar Eqs.~(11) when collisions terms are neglected. It is convenient
to introduce the auxiliary function (see Chuzhoy \& Nusser 2000)
\begin{equation}
\mathcal{F}(\lambda)\equiv \exp\int^\xi{{\rm d}\xi'\over \mathcal{U}-\lambda\xi'}~,
\end{equation}
so that $\mathcal{U}=\lambda\xi+\mathcal{F}/\mathcal{F}'$ and
$\mathcal{U}'=\xi+1-\mathcal{F}\,\mathcal{F}''/(\mathcal{F}')^2$ obtain. Then
from the continuity, mass, energy, and angular momentum equations we find
\begin{eqnarray}
\nonumber &&\mathcal{D}\propto \lambda^{-2}\,\mathcal{F}^{1-3\,\xi}\,\mathcal{F}'~,\\
\nonumber \\
\nonumber &&\mathcal{M}\propto {3\over 2-3\,\xi}\,\mathcal{F}^{2-3\,\xi}~,\\
\\
\nonumber &&\Sigma_r^2\propto \mathcal{F}^{-4\xi}\,(\mathcal{F}')^2~,\\
\nonumber\\
\nonumber &&\Sigma_\theta^2\propto \lambda^{-2}\,\mathcal{F}^{2\,(1-2\xi)}~.
\end{eqnarray}

Rearranging these relations yields
\begin{eqnarray}
\nonumber &&\mathcal{M}\propto {3\over
2-3\,\xi}\,\lambda^2\,\mathcal{D}\,(\mathcal{U}-\lambda\xi)~,\\
\nonumber\\
&&{\Sigma_r^2\over \lambda^4\,\mathcal{D}^2}\propto \mathcal{M}^{\epsilon-2/3}~,\\
\nonumber\\
\nonumber &&\Sigma_\theta^2\,\lambda^2\propto \mathcal{M}^{\epsilon+4/3}~,
\end{eqnarray}
which constitute the `conservation' laws of mass, effective entropy, and
angular momentum. In fact, like in Eqs.~(6), the quantity
$\log[\Sigma_r^2/ \lambda^4\,\mathcal{D}^2] =
\log[(\mathcal{D}\,\lambda^2\,\Sigma_r^2)/(\mathcal{D}\,\lambda^2)^3]$ may be
interpreted as an effective entropy for a one-dimensional fluid with
effective density $\mathcal{D}\, \lambda^2$, effective pressure
$(\mathcal{D}\,\lambda^2)\,\Sigma_r^2$, and microscopic adiabatic index equal
to $3$. Since the mass is a monotonically increasing function of the radius,
it follows that when $\epsilon\la 2/3$ applies such an effective entropy
increases toward smaller $r$; physically, this is interpreted as an efficient
relaxation of the particles during the fast collapse of the halo inner
regions. On the other hand, when $\epsilon\ga 2/3$ applies the effective
entropy grows with $r$; physically, this is interpreted as a progressive
stratification of the particles' orbits (or better, of the orbit apocenters)
during the slow accretion that builds up the halo outskirts (see Bertschinger
1985; also Taylor \& Navarro 2001 for a similar discussion in terms
of the pseudo phase-space density).

When the collision terms are efficient, only the mass conservation still
applies throughout the halo.

\subsection{Asymptotic behaviors}

We now derive analytically the asymptotic behavior of the solutions near
$\lambda\simeq 0$. For the sake of simplicity and with no loss of
generality, we assume the following powerlaw forms of the mean radial
velocity, density, and velocity dispersions\footnote{Given the
constraint $\mathcal{U}(0)=0$ from Eq.~(16), in principle one should write
$\mathcal{U}\sim \lambda^\nu$ with $\nu>0$. However, it can be shown that
Eqs.~(11) do not admit asymptotic solutions for $\nu<1$; hence for $\nu\geq
1$ one can write $\mathcal{U}\sim \mathcal{U}_0\,\lambda$ at the first order.
Moreover, for $\nu\geq 1$ the energy and angular momentum equations imply the
radial and tangential dispersions to scale asymptotically in the same manner,
so that one can write $\Sigma_{r\,,\,\theta}^2\sim \lambda^{\omega}$.}
\begin{eqnarray}
\nonumber &&\mathcal{U}\sim \mathcal{U}_0\,\lambda~,\\
\nonumber\\
&&\mathcal{D}\sim \mathcal{D}_0\, \lambda^{-\gamma}~,\\
\nonumber\\
\nonumber &&\Sigma_{r\,,\,\theta}^2\sim (\Sigma_{r\,,\,\theta}^2)_0\,
\lambda^\omega ~,
\end{eqnarray}
in terms of two exponents $\gamma$ and $\omega$; correspondingly, the mass
behaves as $\mathcal{M}\sim \mathcal{M}_0\, \lambda^{3-\gamma}$ with
$\mathcal{M}_0\equiv 3\,\mathcal{D}_0/(3-\gamma)$, and the pseudo phase-space
density $\mathcal{Q}\equiv \mathcal{D}/\Sigma^3$ follows
$\mathcal{Q}\sim\mathcal{Q}_0\, \lambda^{-\chi}$ with $\mathcal{Q}_0\equiv
\mathcal{D}_0/\Sigma_0^3$ and $\chi=\gamma+3\, \omega/2$.

For such exponents, the continuity and energy equations yield the relations
\begin{eqnarray}
\nonumber && \gamma={-2+3\,\mathcal{U}_0\over \mathcal{U}_0-\xi}~,\\
\\
\nonumber && \omega=-2\,{\xi-1-\mathcal{U}_0\over \mathcal{U}_0-\xi}~.
\end{eqnarray}
In addition, when collision terms are neglected, the momentum equation writes
\begin{equation}
\lambda\,\mathcal{U}_0\,(\mathcal{U}_0-1)+(\Sigma_r^2)_0\,
\lambda^{\omega-1}\,(-\gamma+\omega+2\,\beta_0)+{2\over
3}\,{\mathcal{D}_0\over 3-\gamma}\,\lambda^{1-\gamma}\sim 0~;
\end{equation}
hereafter $\beta_0$ stands for the central value of the anisotropy parameter.

\noindent There are two possibilities for satisfying Eqs.~(21) and (22).

\noindent$\bullet$ First, in Eq.~(22) the exponents of the middle and last
terms are negative and equal, while the coefficient of the middle term must
be negative; then $\omega=-\gamma+2$ applies, and $\gamma>1+\beta_0$ must
hold. Together with Eqs.~(21), these yield $\mathcal{U}_0\simeq 0$ and
\begin{eqnarray}
\nonumber  && \gamma={2\over \xi} = {9\epsilon\over 1+3\epsilon}~,\\
\nonumber\\
&& \omega=2-{2\over \xi} = {2-3\epsilon\over
1+3\epsilon}~,\\
\nonumber\\
\nonumber && \chi=3-{1\over \xi}={3\over 2}\,{2+3\,\epsilon\over
1+3\,\epsilon}~,~~~~~{\rm for}~~~\epsilon \ge {1\over
3}\,{1+\beta_0\over 2-\beta_0}~.
\end{eqnarray}
This case corresponds to the late regime of \emph{slow} accretion of DM
particles onto a preformed halo bulk; in fact, the inner density in physical
units $\rho(t)\propto \lambda^{-\gamma}\, t^{-2}\propto t^{-(2-\gamma\,\xi)}$
behaves as $\rho(t)\propto t^0$ so that it is independent of the time $t$,
i.e., is unaffected by the outskirts growth, consistent with the two-stage
formation picture.

\noindent$\bullet$ Second, in Eq.~(22) the coefficient of the middle term is
zero so that $-\gamma+\omega+2\beta_0=0$ applies, while the difference
$2-\gamma-\omega>0$ between the exponents of the last and middle term must be
positive to imply $\gamma<1+\beta_0$. Together with Eqs.~(21), these
relations yield $\mathcal{U}_0\simeq
4\,[1-6\epsilon+\beta_0\,(1+3\epsilon)]/[9\epsilon\,(2\,\beta_0-5)]$ and
\begin{eqnarray}
\nonumber && \gamma=3\,{3\epsilon+2+2\,\beta_0\over 3\epsilon+7}~,\\
\nonumber\\
&& \omega={3\,(3\epsilon+2)-2\,\beta_0\,(3\epsilon+4)\over
(3\epsilon+7)}~,\\
\nonumber\\
\nonumber && \chi={3\over 2}\,{(3\,\epsilon+2)\,(5-2\,\beta_0)\over
3\,\epsilon+7}~,~~~~~{\rm for}~~~\epsilon <
{1\over 3}\,{1+\beta_0\over 2-\beta_0}~.
\end{eqnarray}
This case corresponds to the early \emph{fast} collapse of the halo bulk; in
fact, the inner density in physical units behaves as $\rho(t)\propto
\lambda^{-\gamma}\, t^{-2}\propto
t^{4\,[1-6\epsilon+\beta_0\,(3\epsilon+1)]/3\epsilon\,(3\,\epsilon+7)}$ so
that it grows with the time $t$. We remark that the asymptotic relation
$\gamma=\omega+2\,\beta_0$ holds like in the solutions of the anisotropic
Jeans equation found by Dehnen \& McLaughlin (2005) and Lapi \& Cavaliere
(2009b). We also stress that the expressions of the asymptotic slopes
in Eqs.~(23) and (24) are continuous at
$\epsilon=(1+\beta_0)/3\,(2-\beta_0)$.

Note that the classic results based on assuming adiabatic invariance
of the radial action may be recovered in the fluid approximation by imposing
the solution to feature for $\lambda\simeq 0$ the additional regularity
condition $t\,\partial_r\, u=\mathcal{U}'(\lambda)\simeq 0$, or equivalently
$\mathcal{U}_0\simeq 0$ (see Teyssier et al. 1997, Subramanian 2000). While
the slope $\gamma$ given by Eq.~(23) for $\epsilon\geq
(1+\beta_0)/3\,(2-\beta_0)$ is consistent with $\mathcal{U}_0\simeq 0$, that
given by Eq.~(24) for $\epsilon< (1+\beta_0)/3\,(2-\beta_0)$ is not as
$\mathcal{U}_0<0$ holds; thus for any $\epsilon$ in the latter range, the
maximal slope $\gamma=1+\beta_0$ consistent with the condition
$\mathcal{U}_0\simeq 0$ is to apply. We give three relevant examples: in a
purely radial collapse with $\beta(r)=\beta_0=1$ the condition
$\mathcal{U}_0\simeq 0$ would imply $\gamma=9\,\epsilon/(1+3\epsilon)$ for
$\epsilon\geq 2/3$ and $\gamma=2$ for any $\epsilon<2/3$, which is the
Fillmore \& Goldreich (1984) result; in an isotropic core with $\beta_0=0$ it
would imply $\gamma=9\,\epsilon/(1+3\epsilon)$ for $\epsilon\geq 1/6$ and
$\gamma=1$ for any $\epsilon< 1/6$; for a slope $\gamma<1$ to hold at the
center it would imply $\beta_0<0$, i.e., prevailing tangential motions
(Subramanian 2000). We remark that although not satisfying the condition
$\mathcal{U}_0\simeq 0$, our asymptotic solution given by Eqs.~(24) is
physical and suited to describe the fast collapse of the halo bulk; to the
best of our knowledge, this was not previously known.

In Fig.~1 we plot against $\epsilon$ the inner asymptotic behaviors given by
Eqs.~(23) and (24) for the slopes of the density $\gamma$ and pseudo
phase-space density $\chi$, for different values of $\beta_0$. We remark that
the full solution attains its asymptotic shape quite slowly, as a result of a
logarithmic convergence; e.g., the solution with $\epsilon=1/6$, that
features a central density slope $-1$, has still a slope around $-1.3$ at
$r\approx 0.1\, r_{-2}$.

When collision terms are efficient, Eqs.~(23) and (24) are still valid, but
the asymptotic behavior
\begin{equation}
\Sigma_r^2-\Sigma_\theta^2\sim \exp\left[-{{3\over
4\,\gamma+3\,\omega-6}\,{\kappa_\star\over \xi-\mathcal{U}_0}}\,
{\mathcal{D}_0\, \mathcal{M}_0\over (\Sigma_{\rm
tot})_0^3}\,\lambda^{3-2\,\gamma-3\,\omega/2}\right]
\end{equation}
applies, and enforces $\beta_0=1-(\Sigma_r^2)_0/(\Sigma_\theta^2)_0\simeq 0$
to hold at the center. Then from Eqs.~(23) and (24) the value
$\epsilon\approx 1/6$ is seen to separate the violent collapse of the inner
region from the calm, inside-out growth of the outskirts. We stress that
dynamical relaxation in the inner region is mainly provided by collective
collisions, whilst in the outskirts it is related to phase-mixing (see also
\S~5).

Note that efficient dynamical relaxation during the fast collapse
stage enforces a vanishing central anisotropy $\beta_0=0$, while making the
inner mass and pseudo phase-space density profiles only weakly dependent on
both the perturbation shape parameter $\epsilon\la 1/6$, and the outer
anisotropy parameter $\beta_+$. In this sense, the inner halo structure turns
out to be approximately \emph{universal}.

\subsection{Numerical solutions}

We solve numerically the system of ordinary differential Eqs.~(11) over the
spatial range $10^{-4}\leq \lambda\leq 1$ with an Adams-Bashford-Moulton
method of variable order, adaptive stepsize, and error control; the location
of the caustic discontinuity is determined with a standard shoot-and-match
technique, by requiring the solution to satisfy the inner constraints
Eqs.~(16), while the jump conditions Eqs.~(15) are applied across the
discontinuity.

As a preliminary check, in Fig.~2 we present the solution for $\epsilon=1$,
$\beta_+=1$, and no collision terms ($\kappa_\star=0$) corresponding to pure
radial infall onto a point-mass perturbation; this is the classic case solved
by Fillmore \& Goldreich (1984) and Bertschinger (1985) basing on
self-similar particle trajectories, and equivalently by Teyssier et al.
(1997) and Subramanian (2000) in the fluid approximation. We recover in
detail the solutions of the latter authors.

Then we focus on the value $\epsilon=1/6$ that corresponds to a spectral
index $n=-2$ typical of a galactic halo. Figs.~3-4-5 refer to $\beta_+=1$,
$\beta_+=0.5$ and $\beta_+=0.25$, respectively, still in the absence of
collisions ($\kappa_\star=0$); these illustrate how the inner density profile
is flattened relative to the purely radial case when non-radial motions are
included. However, as already stressed by Subramanian (2000), Nusser (2001),
and Zukin \& Bertschinger (2010b) such a flattening depends on the
amount of angular momentum assigned at the caustic discontinuity and
on how mass shells are torqued after turnaround, so that producing the
approximately universal shape of the inner density profiles found in
simulations requires fine tuning of a sort in the initial conditions
and/or in the torque mechanism. Moreover, note that in the absence
of collision terms, the anisotropy profile $\beta(r)$ rises inward from the
boundary value $\beta_+$, a behavior at strong variance with what is seen in
numerical simulations.

In Figs.~6-7-8 we retain the values $\epsilon=1/6$ and $\beta_+=0.25$, but
include efficient collision terms; from Fig.~6 to 7 to 8 the strength
parameter $\kappa_\star$ of collisions is increased from $0.01$ to $0.05$ to
$0.1$. The effect of collisions is twofold: the inner slope of the density
profile is now \emph{flattened} to values $\gamma\la 1$, while the anisotropy
parameter $\beta(r)$ is lowered inward to a \emph{vanishing} value
$\beta_0\approx 0$. As collisions become more and more efficient, these
effects occur on wider and wider scales. The same qualitative behavior takes
place for other values of $\epsilon\la 1/6$, as illustrated in Fig.~9 for the
specific case $\epsilon=1/8$, and still with $\beta_+=0.25$ and
$\kappa_\star=0.1$.

In Fig.~10 we plot the position of the caustic discontinuity $\lambda_c$ as a
function of $\epsilon$, for two different values of $\beta_+$, with and
without efficient collision terms. At fixed $\beta_+$ and $\kappa_\star$, it
is seen that $\lambda_c$ increases with $\epsilon$; this is because as
$\epsilon$ grows and the infall rate slows down, the lower infall stress
allows the caustic discontinuity to be located farther out. At fixed
$\epsilon$ and $\kappa_\star$, a lower value of $\beta_+$, corresponding to a
higher angular momentum, yields a larger value of $\lambda_c$; this is
because for particles with higher angular momentum it is harder to penetrate
deep into the halo. Finally, at fixed $\epsilon$ and $\beta_+$, a higher
$\kappa_\star$ yields a smaller $\lambda_c$ since collisions isotropize the
velocity dispersions, and so the particles lose part of their initial angular
momentum and can penetrate deeper into the halo.

We stress that all the above self-similar solutions feature a wide radial
range from the center to about a few $r_{-2}$ where the bulk velocity $u$ is
approximately null. This implies that in the second of Eqs.~(5) the term
$(\partial_t+u\,\partial_r)\, u\simeq 0$ closely vanishes, and the equation
itself reduces to a Jeans form like Eq.~(1). As such it describes a nearly
static equilibrium, endowed with runs of $Q(r)$ and $\beta(r)$ as provided by
the full system of equations.

\subsection{Comparison with $N-$body simulations}

In Figs.~11-14 we compare our self-similar solutions to the outcomes of
state-of-the-art numerical simulations. In all these plots, for the solution
inward of $r_{-2}$ our fiducial values are $\epsilon=1/8$ (as representative
for the range $\epsilon\la 1/6$) and $\kappa_\star=0.1$; in fact, these apply
to the fast inner collapse of galactic halos with effective spectral index
$n=6\epsilon-3\approx -2.25$, when rapid merging of an effective number of
clumps $\mathcal{N}\la 10$ implies \emph{efficient} dynamical relaxation (see
\S~3 and Fig.~A1). For the solution outward of $r_{-2}$, our fiducial values
are $\epsilon=1/2$ (as representative for the range $\epsilon\ga 1/6$) and
$\kappa_\star=0.01$. This is because we expect the outskirts growth to be
dominated by smooth accretion from the tapering perturbation wings with
effective spectral index $n=6\epsilon-3\approx 0$; now the accretion involves
many small clumps with $\mathcal{N}\ga 30$, implying dynamical relaxation to
become \emph{inefficient}. We stress that similar behaviors obtain in the two
radial ranges for reasonable variations of $\epsilon$ and $\kappa_\star$
around our reference values. The overall picture to be compared with real or
simulated data may be obtained on matching these two solutions around
$r_{-2}$.

As to the other parameter $\beta_+$, this is less amenable to physical
pinpointing. We expect $\beta_+ < 1$, i.e., a deviation from purely-radial
collapse, due to asymmetries both in velocity and configuration space. On the
other hand, from the experimentations reported in \S~4.3 we know that in the
presence of $\kappa_\star\ga 0.01$, specific boundary values of $\beta_+$ do
not materially affect the solution, including the run of $\beta(r)$ vanishing
toward the center. Finally, note that around the outer caustic discontinuity
$\beta=1-\Sigma_\theta^2/\Sigma_r^2$ is ill defined anyway, since both the
radial and the tangential dispersions are small upstream (see also end of
this section). Within the above constraints, we adopt the value
$\beta_+=0.25$ both for the inner and outer solutions, because around
$r_{-2}$ this yields similar values of $\beta$, and provides a close match of
the two (already comparable) density slopes.

In Fig.~11 we compare the inner self-similar profiles of density (top panels)
and density slope (bottom panels) with standard fitting formulae to the
outcomes of numerical simulations. Specifically, we illustrate the NFW
profile (dot-dashed line) and the Einasto profile with shape parameter
$\eta=0.17$ (dashed line); in the inner radial range $r\ga 0.04\, r_{-2}$
that is currently accessible to numerical simulations, the former constitutes
a popular description of the virtual data, while the latter has been recently
found to constitute a better functional representation (see \S~1). It is seen
that the inner self-similar solution follows closely the Einasto profile for
$r\ga 0.01\, r_{-2}$, but for smaller radii deviates to attain a central
asymptotic slope $-0.86$ (see \S~4.2 and Fig.~1); in other words, a
steepening of the density profile is predicted for radii $r\la 0.01\, r_{-2}$
relative to the flat Einasto shape. In Fig.~11 we show that a substantially
better representation of the self-similar solution over the whole range
$r<r_{-2}$ is provided by the S\'{e}rsic-Einasto formula (as recalled in
\S~1; cf. also with Graham et al. 2006) with shape parameters $\tau=0.9$ and
$\eta=0.35$. It will be interesting to test such a behavior with numerical
experiments of higher resolutions than presently achieved.

In Fig.~12 we compare the density profile $\rho(r)$ of the self-similar
solutions (thick black solid line) to the outcomes for six different
equilibrium halos extracted from the \textsl{Aquarius} $N-$body simulation
(Navarro et al. 2010; thin colored lines); an Einasto profile with shape
parameter $\eta=0.17$ is also shown for reference (dashed line). The
agreement of the self-similar solutions with the simulation results is
remarkably good for $r\leq$ a few $r_{-2}$. A discrepancy occurs in the outer
regions on approaching the caustic discontinuity; this is expected since our
fluid approximation breaks down there, while the simulated halos themselves
may be out of equilibrium.

In Fig.~13 we compare the self-similar profiles (thick black solid line) of
phase-space density $Q(r)$ to the outcomes from the same six
\textsl{Aquarius} halos plotted in the previous Figure (Ludlow et al. 2010;
thin colored lines); the powerlaw $Q(r)\propto r^{-1.875}$ is also shown for
reference (dashed line). The self-similar profiles agree with the simulation
results as to a close powerlaw shape for $0.04\, r_{-2}\la r\la$ a few
$r_{-2}$. For radii $r\la 0.01\, r_{-2}$ that are not presently probed by
simulations, the self-similar solutions predict a steepening of $Q(r)$ to
attain an asymptotic slope around $2.4$ (as expected on the basis of the
asymptotic analysis of \S~4.2, see Fig.~1; cf. also with Graham et al. 2006);
it will be important to test such a behavior with future simulations of
higher resolution than currently achieved.

The fact that the powerlaw behavior found in simulations is close to
$Q(r)\propto r^{-1.875}$ over an extended range has often been considered
surprising and puzzling; in fact, the value $-1.875$ marks the classic
self-similar solution by Bertschinger (1985) for the purely radial collapse
of a point-mass perturbation. On the other hand, the effective spectral index
$n=3$ corresponding to the Bertschinger's solution constitutes a poor
approximation of the typical overdensity initiating galactic DM halos; in
addition, the resulting inner density profile is quite steeper than found in
galaxy simulations (as recalled in \S~1). Here we see that a powerlaw
behavior close to $Q(r)\propto r^{-1.875}$ in the halo middle is actually
featured by our self-similar solutions with more realistic values $n\approx
-2.25$, when non-radial motions and collision effects are included; we find
that such a powerlaw behavior extends over a considerable radial range around
$r_{-2}$. Inward of this, $Q(r)$ logarithmically steepens while the density
asymptotes to slopes $\la 0.9$; at the other end, beyond $r_{-2}$ toward the
caustic discontinuity $Q(r)$ features an upturn, for which preliminary
evidence has also emerged in recent numerical studies (see Ludlow et al.
2010).

In Fig.~14 we compare the self-similar anisotropy profile (thick black solid
line) with the outcomes from the same six \textsl{Aquarius} halos plotted in
the previous Figures (Ludlow et al. 2010; thin colored lines); the outcome
obtained from the Einasto profile with $\eta=0.17$ and the Hansen \& Moore
(2006) $\beta-\gamma$ relation is also illustrated for reference (dashed
line). Once again, the self-similar solutions agree with the simulation
results (cf. also with the observations by Biviano \& Poggianti 2009). Note
that close to the (outer) caustic discontinuity wide oscillations in the
$\beta=1-\sigma_\theta^2/\sigma_r^2$ parameter are found in simulations, as
expected considering that both the upstream radial and tangential velocity
dispersions are small.

\section{Discussion and conclusions}

Sharing the widespread drive toward understanding the key processes that rule
the dark matter (DM) component of cosmic structures from galaxies to their
systems, we have investigated the detailed structure and evolution of cold DM
halos. To complement the intensive and extensive numerical simulations, here
we have developed advanced models of spherical collapse and accretion in an
expanding Universe, based on solving time-dependent equations for the moments
of the phase-space distribution function in the \emph{fluid} approximation.
Our approach includes non-radial random motions, and most importantly, an
advanced treatment of dynamical \emph{relaxation} effects; thus we provide an
effective description of the halo dynamical evolution toward equilibrium and
of the resulting density and velocity structure. Deferring to \S~5.5 an
overall picture of halo structure and development, we discuss first our
answers to the hot issues on DM dynamics recalled in \S~1.

\subsection{Dynamical relaxation}

Dynamical relaxation involves two mechanisms: \emph{phase-mixing}, due to
spreading of neighboring particle orbits in phase-space; and \emph{violent
relaxation}, due to irregular fluctuations of the gravitational potential.
Phase-mixing can be traced back to the process of \emph{shell crossing} that
underlies the classic self-similar treatments (see Bertschinger 1985; also
Henriksen \& Widrow 1999). In our fluid approach this is expressed by the
left-hand sides of Eqs.~(11), with action primarily confined to a skin region
close to the outer caustic; this circumstance enables one to describe its
effect in terms of boundary jump conditions of Rankine-Hugoniot type. Violent
relaxation, instead, is related to \emph{collective collisions}, expressed by
the terms on the right-hand side of the third and fourth Eqs.~(11), and
active throughout the volume of the halo.

The collision terms have been evaluated quantitatively with the help of a
Fokker-Planck approximation in velocity space, under conditions of a closely
isotropic distribution function (see Appendix A2). To derive the
circularly-smoothed density run $\rho(r)$, the second Eq.~(5) shows that it
is important to understand the behavior of the velocity dispersions
$\sigma_{r\,,\,\theta}^2$; to that effect, we stress that the structure of
our Fokker-Planck coefficients ensures erasure of velocity anisotropy after
$[\partial_t\,(\sigma_{r}^2-\sigma_\theta^2)]_\star\propto
-(\sigma_{r}^2-\sigma_\theta^2)$, while still conserving the overall random
energy after $(\partial_t \sigma_\theta^2)_*=-(\partial_t \sigma_r^2)_*/2$,
see end of \S~2.1.

Anisotropy is initiated at the caustic discontinuity as described by the
parameter $\beta_+$, which specifies the degree of non-radial DM motions
endowed, or acquired during the initial infall. The efficiency of their
erasure is modulated by the strength of the collision terms as expressed by
the parameter $\kappa_\star\la 0.1$; this reflects the amount of
gravitational fluctuations induced by clumpiness in the infalling matter,
well beyond the tiny levels associated to microscopic graininess. Note that
dynamical relaxation acts like an effective `torque' mechanism after
turnaround, in the same vein entertained by Zukin \& Bertschinger (2010a,
2010b); in fact, these authors find results qualitatively similar to ours in
their parametric torque models with decreasing angular momentum and vanishing
central anisotropy.

On these grounds, we have found self-similar solutions for the
spherically-averaged mass density $\rho(r)$, pseudo phase-space density
$Q(r)\equiv \rho/\sigma^3$ and anisotropy parameter $\beta(r)\equiv
1-\sigma_\theta^2/\sigma_r^2$. Overall profiles are obtained on matching (as
discussed in detail in \S~4.4) the self-similar solutions corresponding to
two different portions of the initial DM perturbation, representative of the
fast inner collapse and of the slow outskirt buildup. We have compared these
overall profiles with the outcomes of state-of-the-art $N-$body simulations
throughout the radial range currently probed by the latter, finding a
pleasing agreement.

\subsection{Halo inner regions}

Specifically, in the region inward of a few times $r_{-2}$ (the radius where
the density slope is $-2$) we have found that efficient dynamical relaxation
is the key process to produce the following features: a closely
\emph{universal} density profile in agreement with the $N-$body outcomes,
well represented in terms of the Einasto formula with shape parameter
$\eta\approx 0.17$; a pseudo phase-space density profile with the
\emph{powerlaw} behavior close to $Q(r)\propto r^{-1.9}$; and an anisotropy
profile \emph{decreasing} inward from values around $0.25$ at $r_{-2}$ to
values $\beta\approx 0$ at the center.

In the very central region $r\la 0.04\, r_{-2}$, currently not accessible to
$N-$body simulations, we predict a \emph{steepening} of the mass density
profile relative to the flat Einasto shape, that would imply a vanishing
central slope; the expected asymptotic behavior, though with a
logarithmically slow convergence, reads $\rho(r)\propto r^{-0.86}$, to the
effect that the self-similar density profile is best described by a
S\'ersic-Einasto formula with shape parameters $\tau=0.9$ and $\eta=0.35$. In
parallel, the pseudo phase-space density steepens to approach a central
behavior $Q(r)\propto r^{-2.4}$. Testing these predictions will require
$N-$body simulations of higher resolution than presently achieved.

\subsection{Inner baryonic effects}

Clearly, both our self-similar solutions and $N-$body simulations refer to
pure DM structures. The reader must be aware that in the central regions of
real galaxies small-scale dynamics or energetics related to the astrophysics
of baryons may alter the baseline DM profiles discussed above. These baryonic
processes may reconcile the outcomes of simulations and self-similar models
with the observations of galactic dynamics in spiral galaxies (especially
dwarfs; see Salucci et al. 2007, 2011) and of gravitational lensing in
galaxies (see Brada\v{c} et al. 2009) and galaxy systems (see Zitrin et al.
2011; Newman et al. 2011), which indicate approximately flat density profiles
already inward of $0.1\, r_{-2}$.

For example, flattening of the inner density profile may be caused by
transfer of energy and/or angular momentum from (baryonic and DM) clumps to
the DM field during the galaxy formation process (see El-Zant et al. 2001; Ma
\& Boylan-Kolchin 2004; Tonini et al. 2006). In detail, upon transfer of
tangential random motions from the baryons to an initially isotropic DM
structure, the original density profile $\rho\propto r^{-\gamma}$ in the
inner region is expected to be modified into $\rho\propto
r^{-2\,[\gamma+2\,(2-\gamma)\,\beta_0]/[2+(2-\gamma)\,\beta_0]}$; thus the
final profile is flattened for $\beta_0<0$, down to the point of developing a
core for $\beta_0\la -\tau/2\,(2-\gamma)\approx -0.3$. However, a reliable
assessment of the amount of angular momentum transferred from the baryons to
the DM is still wanting, and would require aimed numerical simulations.

Less agreed processes may affect galactic scales $r\la 10^2$ pc; for example,
at the formation of a spheroid, central starbursts and accretion onto a
supermassive black hole may easily discharge enough energy ($\sim 10^{62}$
erg for a black hole mass $M_\bullet\sim 10^9\, M_\odot$) with sufficient
coupling ($\ga 5\%$) to blow most of the gaseous baryonic mass within
$r_{-2}$ out of the inner gravitational well. This will cause an expansion of
the DM and of the stellar distributions (see Fan et al. 2008, 2010;
Ragone-Figueroa \& Granato 2011), so as to flatten the central DM slope. In
addition, binary black hole dynamics following a substantial merger may eject
on longer timescales formed stars from radii $r\approx 10\, (M_\bullet/10^8\,
M_\odot)^{1/(3-\tau)}$ pc containing an overall mass of a few times the black
hole's, and so may cause the light deficit observed in some galaxy cores (see
Graham 2004; Lauer et al. 2007; Kormendy et al. 2009).

On the other hand, some steepening of the inner density profile may be
induced by any `adiabatic' contraction of the diffuse star-forming baryons
into a disc-like structure, as proposed by Blumenthal et al. (1986) and Mo,
Mao \& White (1998). On the basis of the standard treatments, it is easily
shown that in the inner region an initial powerlaw $\rho\propto r^{-\gamma}$
is turned into $\rho\propto r^{-3/(4-\gamma)}$. However, recent numerical
experiments (see Abadi et al. 2010) suggest that the classic treatment of
adiabatic contraction is likely to be extreme; actually, in the inner region
the contraction may be inefficient and the density slope hardly modified.

All these issues are beyond the scope of the present paper.

\subsection{Halo outskirts}

At the other end, in the region outward of a few times $r_{-2}$ we find that
self-similar solutions with larger $\epsilon\ga 1/2$ and \emph{less}
efficient dynamical relaxation apply, consistent with the outcomes of
numerical simulations. This is expected, since the outskirts are built up at
later times by smoother (i.e., less clumpy) accretion fed on the tapering
wings of a DM perturbation. We note that, as highlighted by the scatter in
the $N-$body results, the outskirts structure is subject to a large variance,
related to the detailed growth histories and to environmental conditions
wherefrom the infall takes place.

Relatedly, for redshifts $z\la 0.5$ the accelerating cosmology slows down the
time dependence $D(t)\propto t^{d}$ of the growth factor from $d=2/3$ to
$d\approx 1/2$. A strictly self-similar solution cannot be obtained in such
conditions, but the overall trend can be captured from noting that the
accretion rate scales as $\dot M/M=d/\epsilon\, t$; thus a lower $d$ is
equivalent to higher effective values of $\epsilon$, which result in steeper
density profiles into the outskirts. Moreover, in a flat accelerating
Universe, shells will be able to turnaround and collapse only if their
initial position $r_i$ is inside a critical radius $r_{\Lambda}$ defined by
$\delta M/M (<r_\Lambda)\equiv 3/2\,(2\, \Omega_\Lambda)^{1/3}$ in terms of
the dark energy density parameter $\Omega_\Lambda\approx 0.7$; Subramanian et
al. (2000) have shown that the resulting outer density profile is
considerably steep, featuring a cutoff toward $r_\Lambda$.

Our self-similar solutions concur with numerical simulations in providing a
firm basis and in clarifying the limitations for the simple dynamical models
based on the Jeans equation and on a consistent powerlaw shape for
$Q(r)\propto r^{-1.9}$, see \S~1 for details. We find that such models are
reliable in the significant range $0.01\, r_{-2}\la r\la$ a few $r_{-2}$;
outside of this, in the outskirts $Q(r)$ flattens appreciably and is liable
to cosmogonical variance, while in the inner regions it steepens, though
logarithmically. It may be interesting to introduce in the dynamical models
this articulated behavior of $Q(r)$, to investigate how the Jeans profiles
are affected (taking up from the work by Lapi \& Cavaliere 2009b).

\subsection{An overall picture of halo structure and development}

Finally, we stress the \emph{link} between the halo structure and its
two-stage growth history from the vantage point of our self-similar
solutions, to provide the following overall picture (illustrated in the
schematics of Fig.~15).

The halo formation history starts up with the fast collapse of a few merging
clumps, that simulations indicate to be the main contributors to the inner
halo mass (see Fakhouri et al. 2010; Genel et al. 2010; Wang et al. 2011);
such a strong clumpiness enforces efficient dynamical relaxation, the key
process to produce an approximately \emph{universal} shape of the inner
density profile, and an isotropic inner halo structure.

As discussed in \S~3, the typical halo mass and radius scale as $M\propto
t^{2/3\epsilon}$ and $R\propto t^{2/3+2/9\epsilon}$ in terms of the
perturbation shape parameter $\epsilon$, or of the corresponding effective
index of the perturbation spectrum $n\approx 6\,\epsilon-3$. Thus the fast
collapse of the \emph{inner} region affected by violent relaxation is seen to
correspond to $\epsilon\la 1/6$ or $n\la -2$, since for these values the
depth of the gravitational well expressed by $v_c\sim (G\, M/R)^{1/2}\propto
t^{2/9\epsilon-1/3}$ grows faster than the cosmic time. These values are also
consistent with the asymptotic analysis of \S~4.2. We also note that for
$n\la 1$ or $\epsilon\la 2/3$ the typical specific energy $v_c^2\propto
M^{2/3-\epsilon}$ increases with mass, so that the inner region provides an
environment conducive to eventual melting of the accreted massive clumps.
This validates spherically-averaged densities, and is consistent with the
findings from numerical simulations that the residual mass fraction still
locked into clumps is limited in the inner regions (see Springel et al. 2008,
their Figs.~11-12). Note that the melting of clumps due to tidal disruption
and dynamical friction may be even more efficient in a warm DM framework
(e.g., Col\'{i}n et al. 2000).

Subsequently, a stage of calmer and smoother accretion ensues; this is slower
relative to the cosmic expansion when $n\ga 1$ or $\epsilon\ga 2/3$, since
the growth rate scales as $M/t\propto t^{2/3\epsilon-1}$. During this stage,
simulations indicate that most of the mass is gathered about equally from
minor mergers of many small clumps and from truly diffuse accretion (see
above references). Then the \emph{outskirts} build up from the inside-out,
while the preformed inner region is nearly unaffected; this is because the
smoothness of the accretion makes dynamical relaxation inefficient and only
mild phase-mixing occurs at the outer caustic (see \S~3 and 4.2). Meanwhile,
cosmological variance and non-radial motions related to the halos' specific
growth history or environmental conditions become relevant in shaping the
halo outskirts, in ways expected to be far from universal.

We envisage such a two-stage framework of halo growth to be of much relevance
for the astrophysics of galaxy clusters, especially in their outskirts as
discussed by Lapi et al. (2010), and for galaxy/star formation theories;
e.g., in the latter context it is at the hearth of the two-phase galaxy
formation picture first proposed by Cook et al. (2009). This envisages that
the two modes of halo growth drive two distinct modes for the evolution of
baryonic matter, with the development of the spheroidal component of galaxies
taking place mainly during the early fast collapse, and that of the disc
component during the late slow accretion.

\begin{acknowledgements}
Work supported in part by ASI, INAF and MIUR. We thank our referee for
helpful comments and suggestions. We acknowledge useful discussions with L.
Danese, G. De Zotti, G.L. Granato, and P. Salucci. AL thanks SISSA for warm
hospitality.
\end{acknowledgements}

\begin{appendix}

\section{The collision terms}

In this Appendix we evaluate the collision terms to be inserted in Eqs.~(5)
and (11). We take up and adapt to the present problem the approach developed
by Larson (1969, 1970) in the context of stellar clusters.

\subsection{The velocity distribution function}

First of all, we specify the form of the velocity distribution function. We
introduce spherical polar coordinate in velocity space, and write the
velocity components as
\begin{eqnarray}
\nonumber && v_r-u=\nu\,\mu~,\\
\\
\nonumber && v_\theta^2+v_\phi^2=\nu^2\,(1-\mu^2)~,
\end{eqnarray}
in terms of the random velocity vector $\nu$ and the cosine $\mu$ of the
angle between this vector and the radial direction.

We assume that the normalized distribution $f(\nu,\mu)$ of random velocities
is approximately isotropic, and expand it in Legendre polynomials up to the
second order
\begin{equation}
f(\nu,\mu)\simeq f_0(\nu)\,P_0(\mu)+f_1(\nu)\,P_1(\mu)+f_2(\nu)\,P_2(\mu)~;
\end{equation}
the coefficients $f_1$ and $f_2$ are assumed to be small corrections to the
leading term $f_0(\nu)$ provided for definiteness by a Maxwellian-like
distribution.

The normalization conditions for $f$ require that
\begin{equation}
\int{\rm d}^3\nu~ f=1~~~~~~,~~~~~~\int{\rm d}^3\nu~ \nu\, f=0~
\end{equation}
hold, with $\int{\rm d}^3\nu=2\pi\,\int_0^{\infty}{\rm d}\nu\,
\nu^2\int_{-1}^1{\rm d}\mu$ being the volume element. These imply that the
terms in the expansion Eqs.~(A2) must depend only on the combinations $\nu^2$
or $\nu\, \mu$; thus we base on the expressions
\begin{eqnarray}
\nonumber && f_0(\nu)={1\over (2\pi\,\sigma^2)^{3/2}}\,\exp\left(-{\nu^2\over
2\sigma^2}\right)~~~~~~~~~P_0(\mu)=1~,\\
\nonumber\\
&& f_1(\nu)=c_1\,\nu\,f_0(\nu)~~~~~~~~~~~~~~~~~~~~~~~~~~~~~P_1(\mu)=\mu~,\\
\nonumber\\
\nonumber && f_2(\nu)=c_2\,\nu^2\,f_0(\nu)~~~~~~~~~~~~~~~~~~~~~~~~~~~P_2(\mu)={3\,\mu^2-1\over
2}~,
\end{eqnarray}
where $c_1$ and $c_2$ are constants to be determined.

As discussed in \S~2, to close the moment equations we are interested in a
distribution with zero skewness, which gives $c_1=0$. On the other hand,
$c_2$ can be determined by considering the second-order moments of the
distribution, i.e.,
\begin{eqnarray}
\nonumber && \sigma_r^2=\int{\rm d}^3\nu~ \nu^2\,\mu^2\, f=\sigma^2+2\,c_2\,\sigma^4~,\\
\\
\nonumber && \sigma_\theta^2=\int{\rm d}^3\nu~\nu^2\,(1-\mu^2)\,
f=\sigma^2-c_2\,\sigma^4~;
\end{eqnarray}
since $\sigma_{\rm tot}^2\equiv\sigma_r^2+2\,\sigma_\theta^2=3\,\sigma^2$ and
$\sigma_r^2-\sigma_\theta^2=3\, c_2\,\sigma^4$ hold, we find
\begin{equation}
c_2=3\, {\sigma_r^2-\sigma_\theta^2\over \sigma_{\rm tot}^4}~.
\end{equation}

\subsection{Fokker-Planck approximation}

Now we turn to the form of the collision terms. In the Fokker-Planck
approximation, these are written as
\begin{equation}
\Gamma_\star^{-1}\, (\partial_t\, f)_\star=-\sum_i\partial_{v_i}\,(f\,\partial_{v^i}\,h)+{1\over
2}\,\sum_{i,j}\,\partial_{v_i\,v_j}^2\,(f\,\partial_{v^i\,v^j}\, g)
\end{equation}
in terms of
\begin{eqnarray}
\nonumber && h=2\, \int{\rm d}^3\nu'~{f(\nu')\over |\nu-\nu'|}~,\\
\\
\nonumber && g=2\, \int{\rm d}^3\nu'~f(\nu')\,|\nu-\nu'|~.
\end{eqnarray}
The quantity $\Gamma_\star\equiv 4\pi\,G^2\, m\, \rho\, \log\Lambda$
represents the strength parameter of the collective collisions, and depends
on the clump mass $m$ as $m\,\log M/m$; here the standard `Coulomb' logarithm
$\log \Lambda$ has been conveniently written in terms of the ratio $M/m$
between the halo and the clump masses (see Boylan-Kolchin et al. 2008). In
fact, numerical simulations indicate that the clumps (technically,
`subhalos') in a halo are distributed close to $m^{-2}$ (see Springel et al.
2008); averaging $\Gamma_\star$ over such a distribution turns out to be
closely equivalent to redefine $m$ as the average clump mass (see Ciotti
2010).

In axial symmetry, the above terms $h$ and $g$ can be simplified considerably
(see Rosenbluth et al. 1956), and to the lowest-order one obtains
\begin{eqnarray}
\nonumber && h\simeq 8\pi\,\left[\int_0^\nu{\rm d}\nu'~{\nu'^2\over
\nu}\,f_0(\nu')+\int_\nu^\infty{\rm
d}\nu'~\nu'\,f_0(\nu')\right]~,\\
\\
\nonumber && g\simeq 4\pi\,\left[\nu\,\int_0^\nu{\rm d}\nu'~\nu'^2\, \left(1+{\nu'^2\over
3\,\nu^2}\right)\,f_0(\nu')+\int_\nu^\infty{\rm d}\nu'~\nu'^3\,\left(1+{\nu^2\over
3\,\nu'^2}\right)\,f_0(\nu')\right]~;
\end{eqnarray}
correspondingly, the Fokker-Planck equation simplifies to
\begin{eqnarray}
\nonumber \Gamma_\star^{-1}\, (\partial_t\, f)_\star =&-& {1\over
\nu^2}\,\partial_\nu\,(f\,\nu^2\,\partial_\nu\,h)+{1\over
2\nu^2}\,\partial_{\nu\nu}^2\,(f\,\nu^2\,\partial_{\nu\nu}^2\, g)+{1\over
2\nu^2}\,\partial_{\mu\mu}^2\,\left[f\,{1-\mu^2\over \nu}\,\partial_\nu\, g\,\right]+\\
\\
\nonumber  &-& {1\over \nu^2}\,\partial_\nu\,(f\,\partial_\nu\,g)+{1\over
\nu^2}\,\partial_\mu\,\left(f\,{\mu\over \nu}\,\partial_\nu\, g\right)~.
\end{eqnarray}

Introducing the relevant integrals
\begin{eqnarray}
\nonumber && \mathcal{I}=\int_0^\nu{\rm d}\nu'~\nu'^2\,f_0(\nu') ~,\\
\nonumber\\
&& \mathcal{J}=\int_0^\nu{\rm d}\nu'~\nu'^4\,f_0(\nu')~,\\
\nonumber\\
\nonumber && \mathcal{K}=\int_\nu^\infty{\rm d}\nu'~\nu'\,f_0(\nu')~,
\end{eqnarray}
after some manipulation one finds that
\begin{eqnarray}
\nonumber && \partial_\nu\, h=-{8\pi\over \nu^2}\, \mathcal{I}~~~~~~\partial_{\nu\nu}^2\,
h=-8\pi\,f_0~,\\
\\
\nonumber && \partial_\nu\, g=4\pi\,\left(\mathcal{I}-{\mathcal{J}\over 3\nu^2}+{2\over
3}\,\nu\,\mathcal{K}\right)~~~~~~\partial_{\nu\nu}^2\, g={8\pi\over
3}\,\left(\mathcal{K}+{\mathcal{J}\over
\nu^3}\right)~~~~~~\partial_{\nu\nu\nu}^3\,g=-8\pi\,{\mathcal{J}\over \nu^4}~.
\end{eqnarray}

Corresponding to the Legendre expansion for $f$ in Eq.~(A2), it is convenient
to perform the same for $(\partial_t\, f)_\star$, to read
\begin{equation}
(\partial_t\, f)_\star\simeq (\partial_t\, f_0)_\star\, P_0(\mu)+(\partial_t\,
f_2)_\star\,P_2(\mu)~.
\end{equation}
Then the Fokker-Planck equation yields
\begin{eqnarray}
\nonumber && \Gamma_\star^{-1}\,(\partial_t\,f_0)_\star\simeq {4\pi\over \nu^2}\,{{\rm d}\over {\rm
d}\nu}\left[f_0\,
\mathcal{I}+{1\over 3\nu}\,(\mathcal{J}+\nu^3\,\mathcal{K})\,{{\rm d}\over {\rm
d}\nu}f_0\right]~,\\
\\
\nonumber && \Gamma_\star^{-1}\,(\partial_t\,f_2)_\star\simeq {4\pi\over \nu^2}\,{{\rm d}\over {\rm
d}\nu}\left[f_2\,
\mathcal{I}+{1\over 3\nu}\,(\mathcal{J}+\nu^3\,\mathcal{K})\,{{\rm d}\over {\rm
d}\nu}f_2\right]-{4\pi\over
\nu^5}\,(3\nu^2\,\mathcal{I}+2\nu^3\,\mathcal{K}-\mathcal{J})\,f_2~.
\end{eqnarray}

Taking now the second-order moments leads to the collision terms
\begin{eqnarray}
\nonumber && (\partial_t\, \sigma_r^2)_\star=8\pi^2\,\Gamma_\star\,\int_0^\infty{\rm
d}\nu~\left[{4\over
3}\,\nu\,(\nu\,\mathcal{K}-\mathcal{I})\,f_0-{4\over
15}\,\left(5\nu\,\mathcal{I}-{\mathcal{J}\over
\nu}\right)\,f_2\right]~,\\
\\
\nonumber && (\partial_t\, \sigma_\theta^2)_\star=8\pi^2\,\Gamma_\star\,\int_0^\infty{\rm
d}\nu~\left[{4\over
3}\,\nu\,(\nu\,\mathcal{K}-\mathcal{I})\,f_0+{2\over
15}\,\left(5\nu\,\mathcal{I}-{\mathcal{J}\over
\nu}\right)\,f_2\right]~.
\end{eqnarray}

With the adopted form of $f$ discussed in \S~A1, we find
\begin{equation}
\int_0^\infty{\rm
d}\nu~\nu\,(\nu\,\mathcal{K}-\mathcal{I})\,f_0=0~~~~~~,~~~~~~\int_0^\infty{\rm
d}\nu~\left(5\nu\,\mathcal{I}-{\mathcal{J}\over
\nu}\right)\,f_2={c_2\,\sigma\over 2\,\pi^{5/2}}~,
\end{equation}
so that the collision terms finally read
\begin{eqnarray}
\nonumber && (\partial_t\, \sigma_r^2)_\star=-{64\,\sqrt{3\, \pi}\over
15}\,G^2\,m\,\rho\,\log\Lambda\,{\sigma_r^2-\sigma_\theta^2\over \sigma_{\rm tot}^3}~,\\
\\
\nonumber && (\partial_t\, \sigma_\theta^2)_\star=+{32\,\sqrt{3\,\pi}\over
15}\,G^2\,m\,\rho\,\log\Lambda\,{\sigma_r^2-\sigma_\theta^2\over \sigma_{\rm
tot}^3}~.
\end{eqnarray}

With the effective number of clumps $\mathcal{N}\equiv M/m$ approximately
constant, the above collision terms do not break self-similarity, and can be
written as
\begin{eqnarray}
\nonumber && (\Sigma_r^2)_\star'=-\kappa_\star\,
{\mathcal{D}\,\mathcal{M}\over \Sigma_{\rm tot}^3}\,
(\Sigma_r^2-\Sigma_\theta^2)~,\\
\\
\nonumber && (\Sigma_\theta^2)_\star'=+{\kappa_\star\over 2}\,
{\mathcal{D}\,\mathcal{M}\over \Sigma_{\rm tot}^3}\,
(\Sigma_r^2-\Sigma_\theta^2)~.
\end{eqnarray}
in terms of the self-similar variables defined in Eqs.~(10) of the main text.
The strength parameter of dynamical relaxation
\begin{equation}
\kappa_\star\equiv {64\,\sqrt{3\pi}\over 405}\,{\log \mathcal{N}\over
\mathcal{N}}\approx 0.5\, \,{\log \mathcal{N}\over
\mathcal{N}}
\end{equation}
depends mainly on the clumpiness $1/\mathcal{N}$ of the infalling
matter. In Fig.~A1 we illustrate $\kappa_\star$ as a function of
$\mathcal{N}$, and highlight the regimes corresponding to the early fast
collapse and to the late slow accretion. In more detail, during the early
fast collapse of the halo inner region, a limited effective number of clumps
$\mathcal{N}\la 10$ is expected to enforce efficient dynamical relaxation
with $\kappa_\star\approx 0.1$, while during the late slow accretion
dominating the growth of the halo outskirts, many small infalling clumps with
$\mathcal{N}\ga 30$ make dynamical relaxation much less efficient with
$\kappa_\star\la 0.01$.

\end{appendix}

\clearpage
\begin{figure}
\epsscale{0.7}\plotone{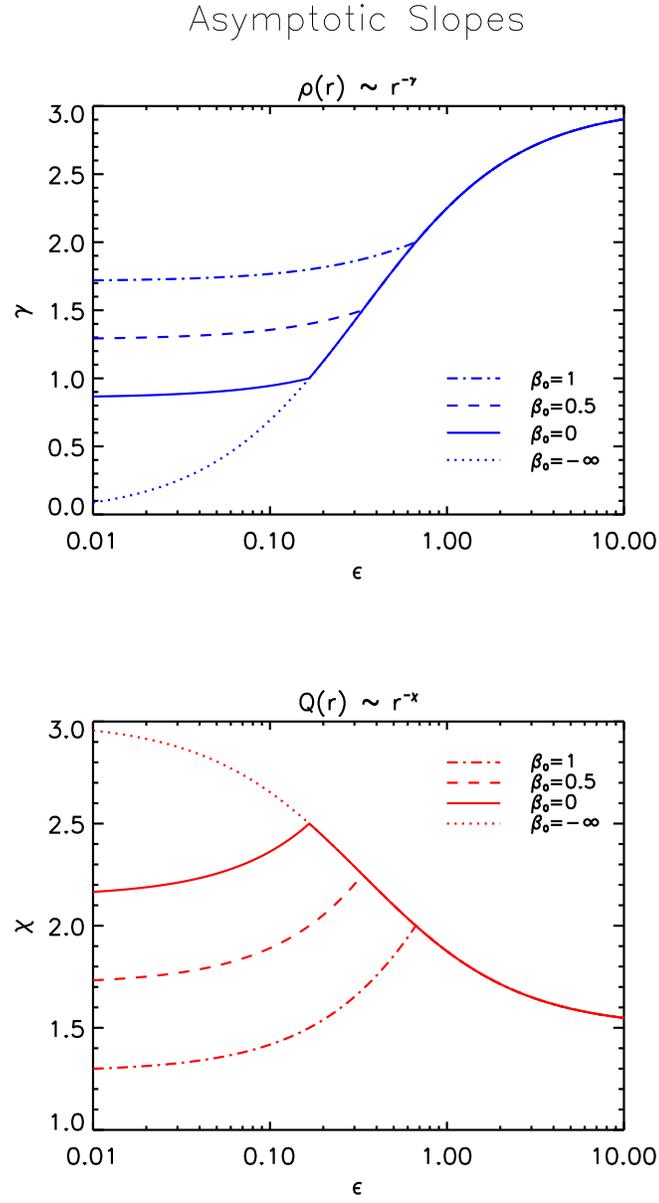}
\caption{Asymptotic inner slopes of density and (pseudo) phase-space density
as a function of $\epsilon$, for different values of the central anisotropy
parameter. We recall from \S~2.1 that efficient dynamical relaxation enforces
$\beta_0=0$.}
\end{figure}

\clearpage
\begin{figure}
\epsscale{1}\plotone{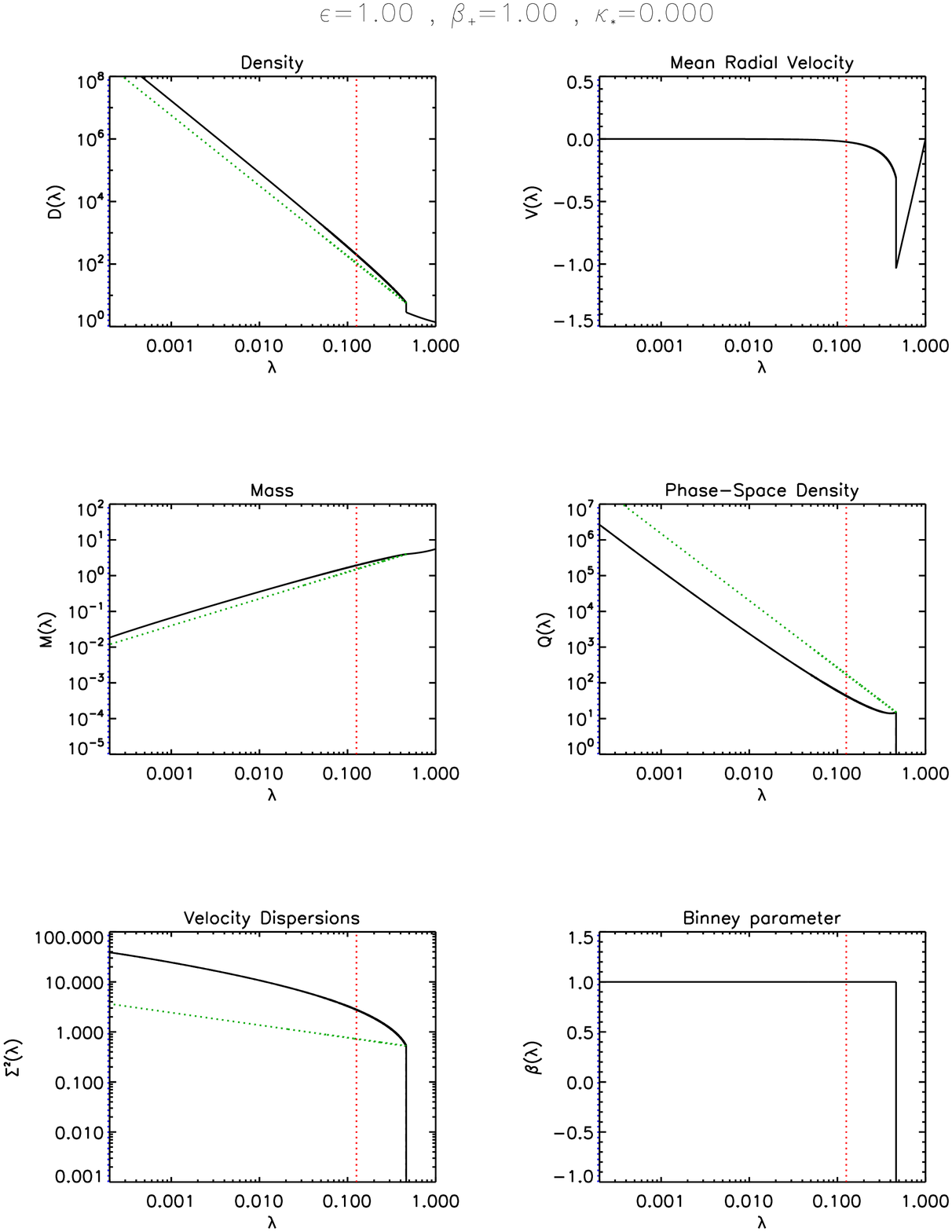}
\caption{The self-similar solution for $\epsilon=1$ and $\beta_+=1$
corresponding to purely radial orbits, in the absence of collisions
($\kappa_\star=0$); this is just the case considered by Bertschinger (1985).
Green dotted lines illlustrate the expected asymptotic behaviors (see
\S~4.2). Vertical red lines mark the radius $R_{200}$ where the density
amounts to $200$ times the background's. Same linestyles are adopted in the
following figures.}
\end{figure}

\clearpage
\begin{figure}
\epsscale{1}\plotone{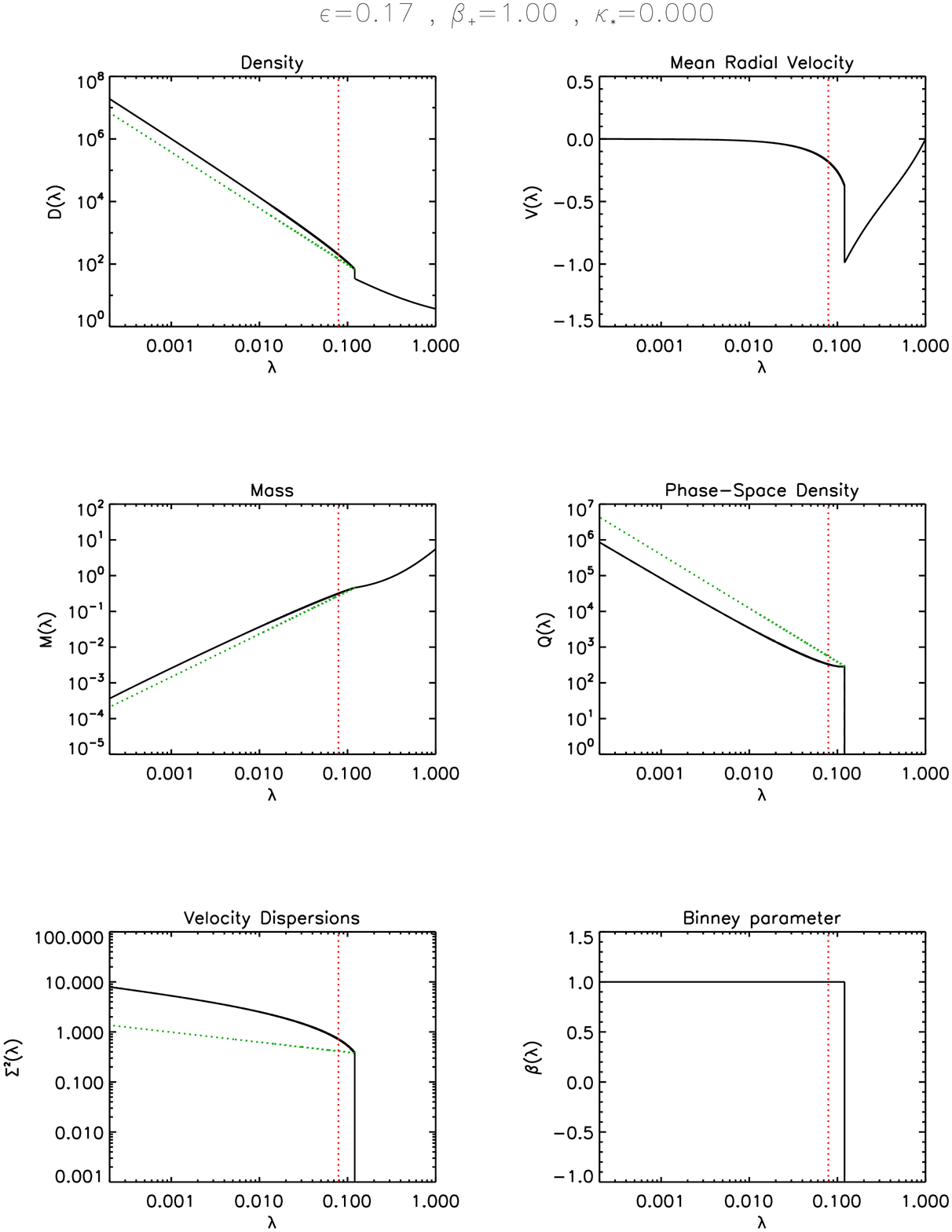}
\caption{The self-similar solution for $\epsilon=1/6$ and $\beta_+=1$
corresponding to purely radial orbits, in the absence of collisions
($\kappa_\star=0$).}
\end{figure}

\clearpage
\begin{figure}
\epsscale{1}\plotone{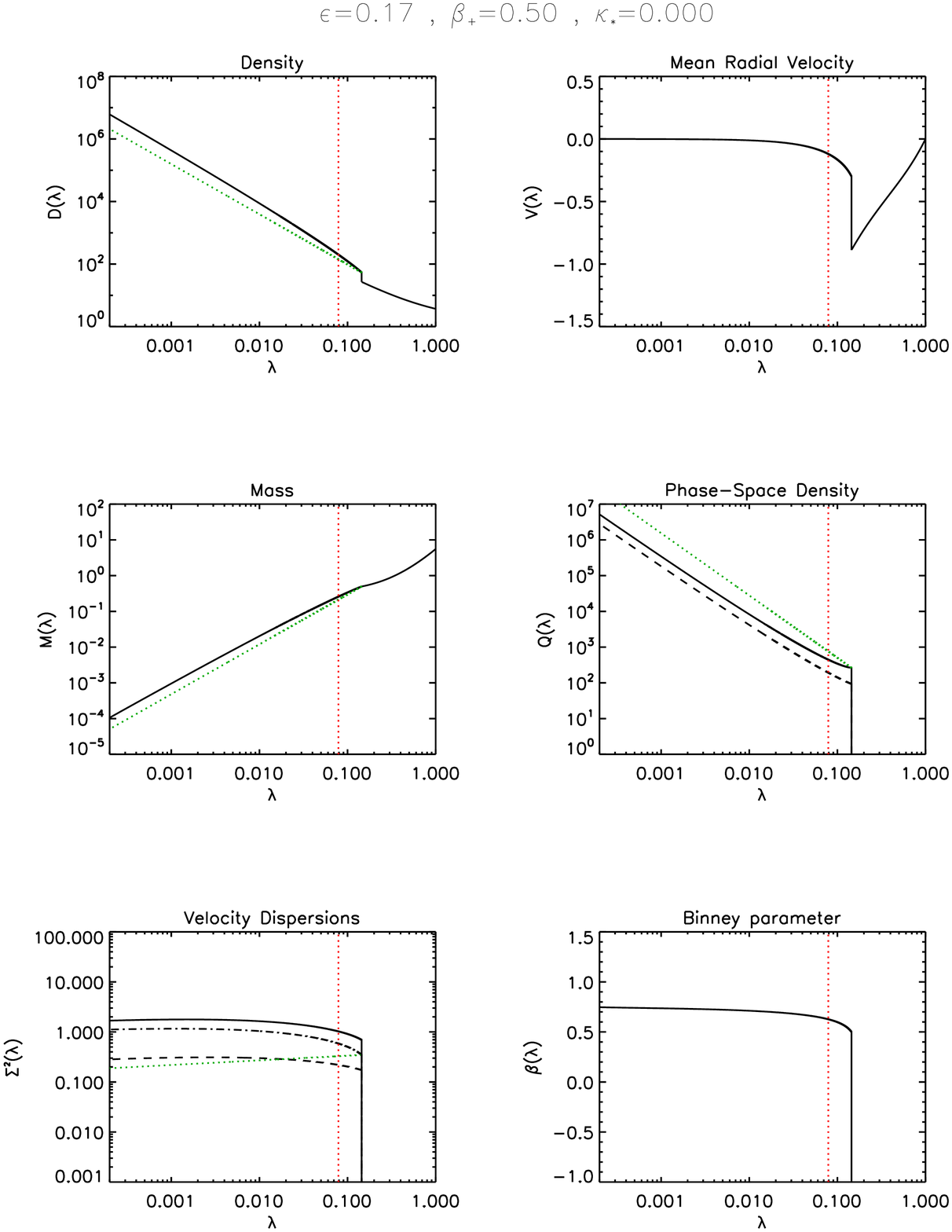}
\caption{The self-similar solution for $\epsilon=1/6$ and $\beta_+=0.5$ in the
absence of collisions $\kappa_\star=0$. In the middle-right panel, solid line
refers to $Q(r)$ defined in terms of the total velocity dispersion, while
dashed line refers to $Q(r)$ defined in terms of the radial dispersion. In
the bottom-left panel, solid line refers to the total velocity dispersion,
dot-dashed line to the radial component and dashed line to the tangential
component.}
\end{figure}

\clearpage
\begin{figure}
\epsscale{1}\plotone{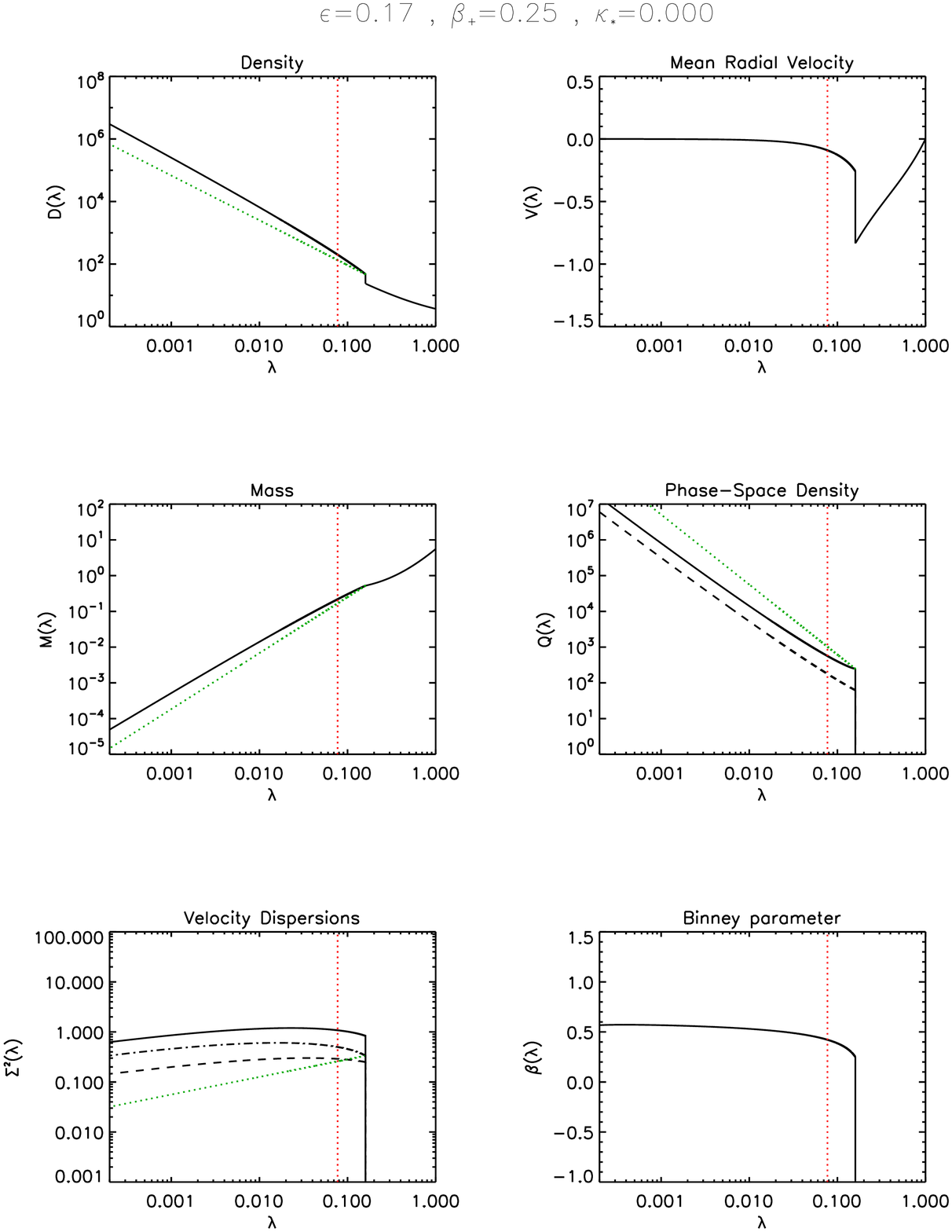}
\caption{Self-similar solution for $\epsilon=1/6$ and
$\beta_+=0.25$, in the absence of collisions $\kappa_\star=0$.}
\end{figure}

\clearpage
\begin{figure}
\epsscale{1}\plotone{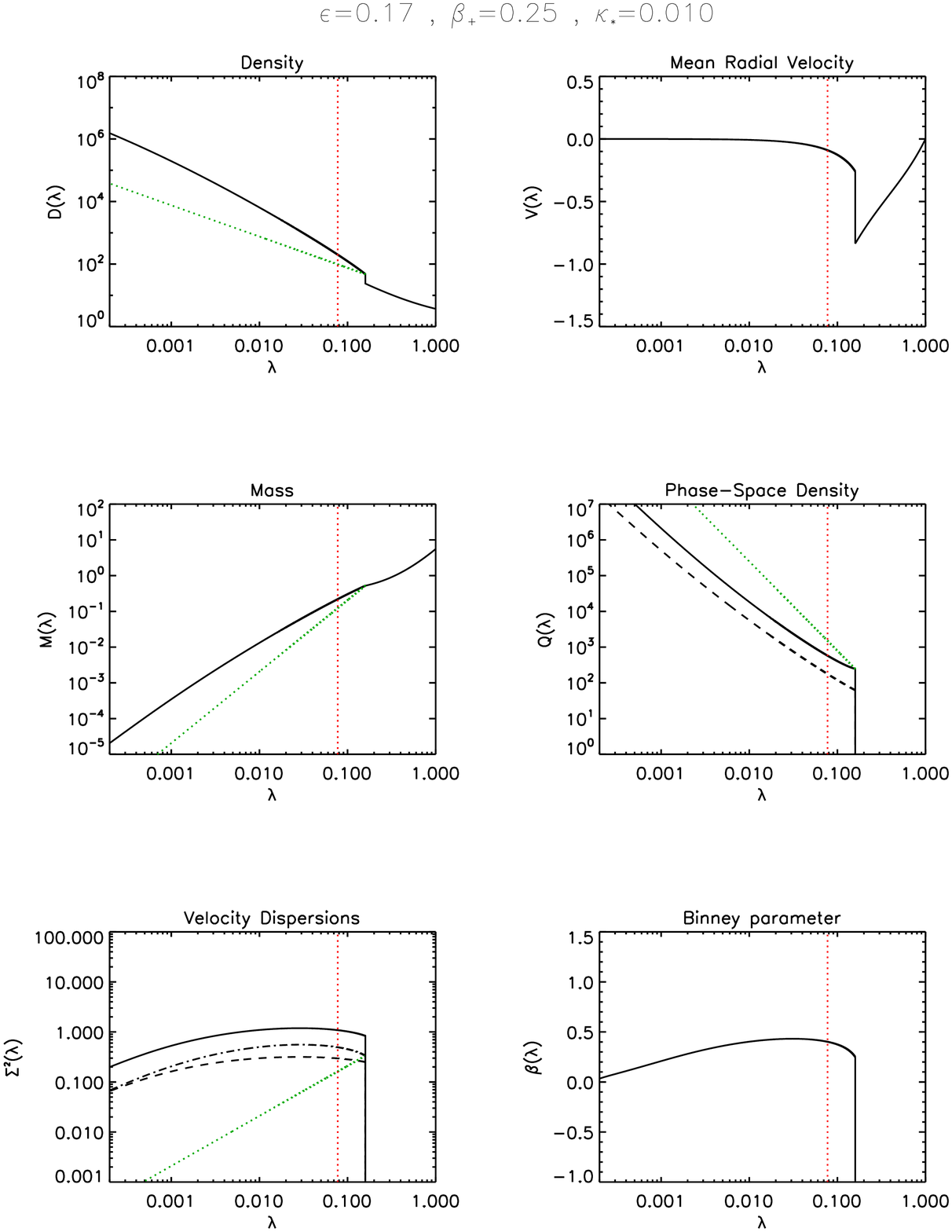}
\caption{Self-similar solution for $\epsilon=1/6$ and $\beta_+=0.25$, in
presence of collisions with strength parameter $\kappa_\star=0.01$.}
\end{figure}

\clearpage
\begin{figure}
\epsscale{1}\plotone{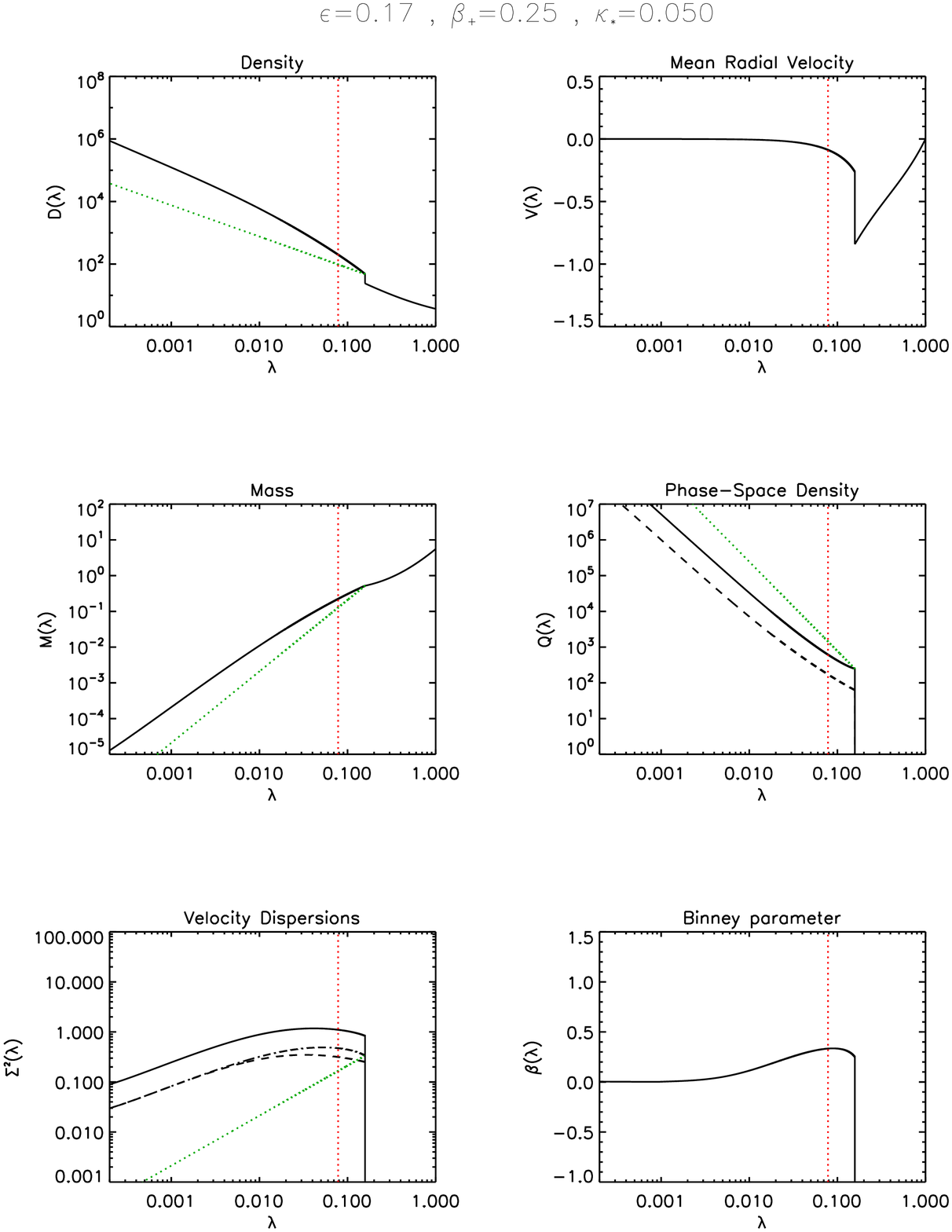}
\caption{Self-similar solution for $\epsilon=1/6$ and $\beta_+=0.25$, in
presence of collisions with strength parameter $\kappa_\star=0.05$.}
\end{figure}

\clearpage
\begin{figure}
\epsscale{1}\plotone{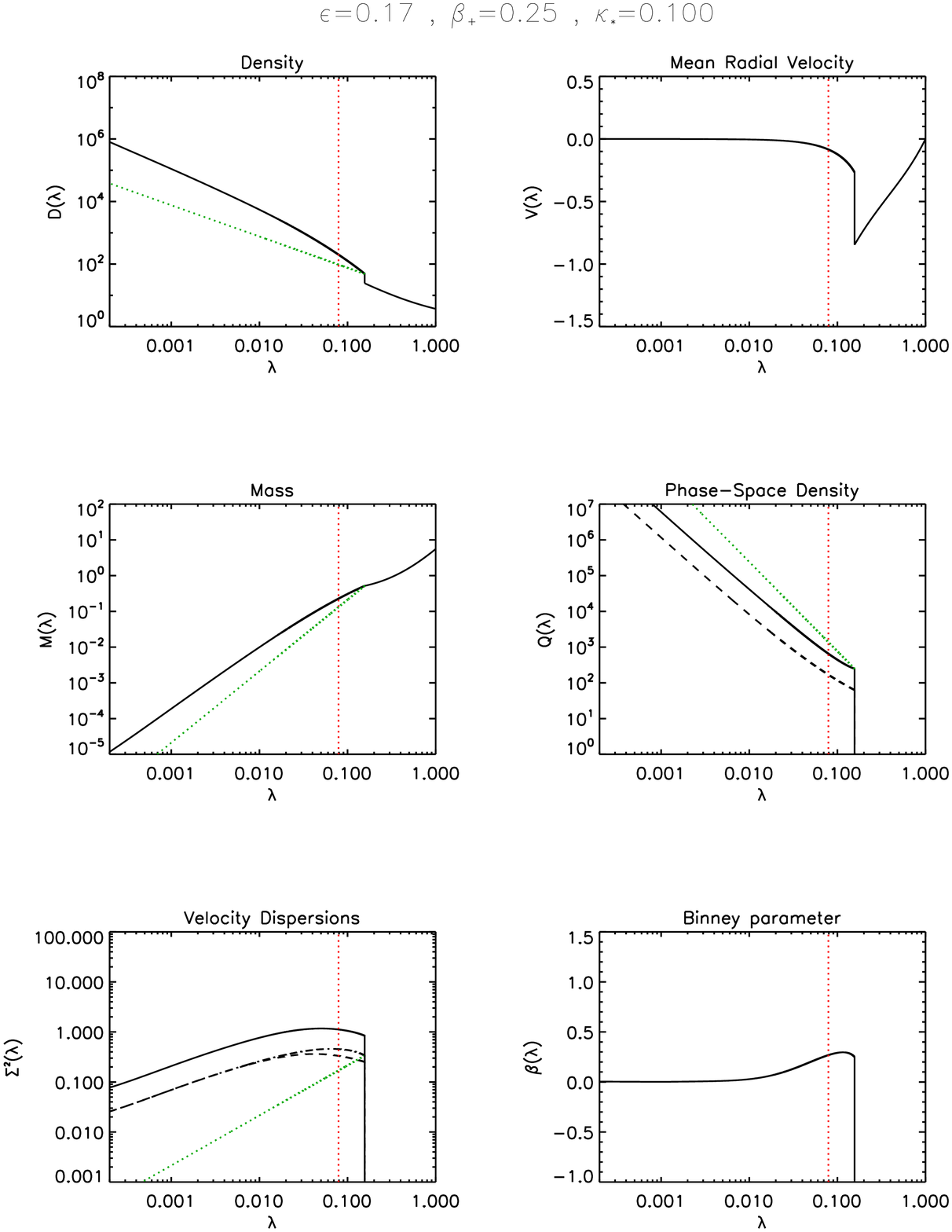}
\caption{Self-similar solution for $\epsilon=1/6$ and $\beta_+=0.25$, in
presence of collisions with strength $\kappa_\star=0.1$.}
\end{figure}

\clearpage
\begin{figure}
\epsscale{1}\plotone{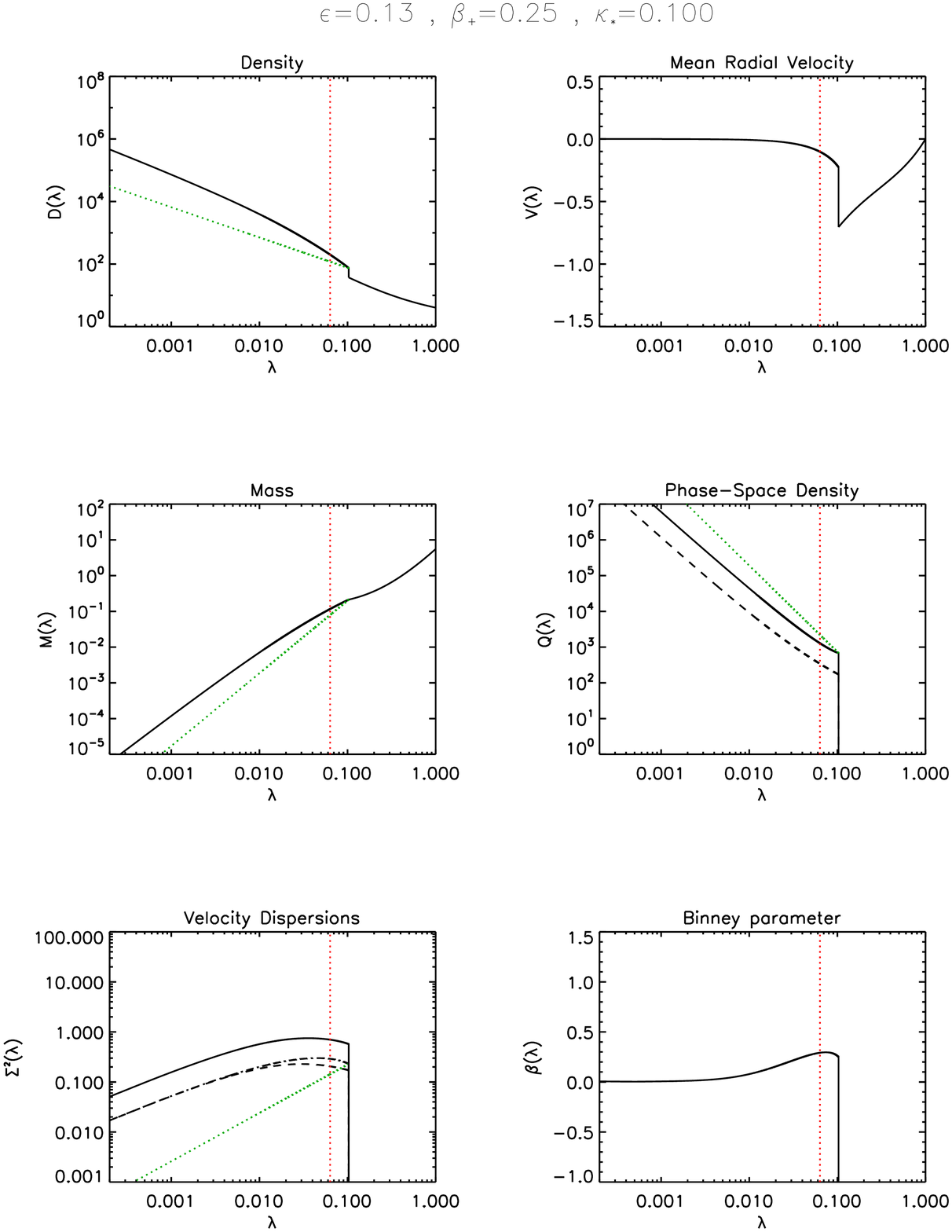} \caption{Self-similar solution for
$\epsilon=1/8$ and $\beta_+=0.25$, in presence of collisions with
strength $\kappa_\star=0.1$.}
\end{figure}

\clearpage
\begin{figure}
\epsscale{0.8}\plotone{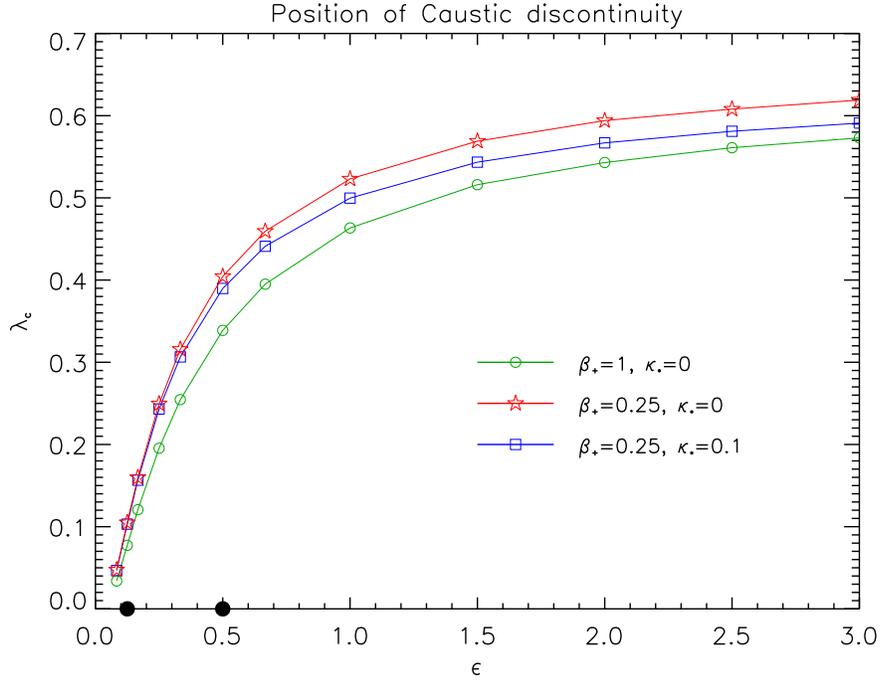} \caption{Position of the caustic
discontinuity as a function of $\epsilon$, for two values of $\beta_+$ and
$\kappa_\star$; these parameters only mildly affect the caustic position at
given $\epsilon$. The curves have been computed at discrete values of
$\epsilon$, as highlighted with the open symbols; the specific values
$\epsilon=1/8$ and $1/2$ used in the next figures are marked by filled dots.
Note that for values $\epsilon\ga 1/2$ the caustic position are close to the
standard virial radius $R_{200}\approx R_{\rm ta}/2$ recalled in \S~3.}
\end{figure}

\clearpage
\begin{figure}
\epsscale{1}\plotone{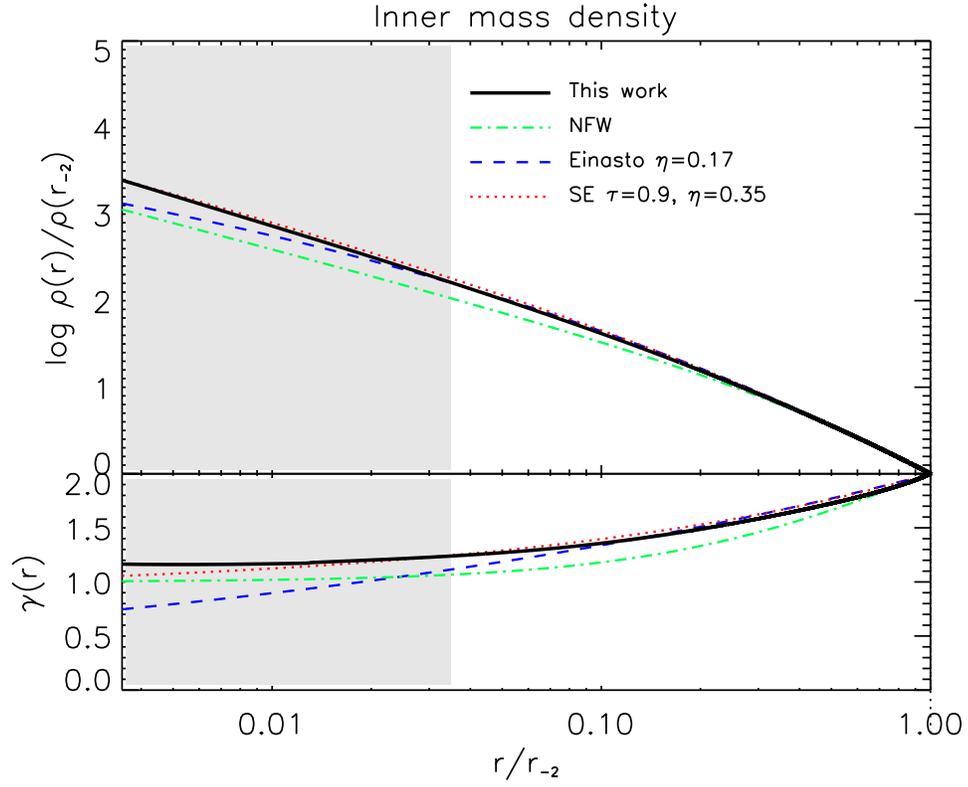}
\caption{Comparison of the inner self-similar density profile (top) and
density slope (bottom) with standard fitting functions to the equilibrium
outcomes of numerical simulations. Solid line refers to the self-similar
solution with parameters $\epsilon=1/8$, $\beta_+=0.25$, $\kappa_\star=0.1$,
dot-dashed line refers to the NFW profile, dashed line refers to the Einasto
profile with shape parameter $\eta=0.17$, dotted line refers to the
S\'{e}rsic-Einasto profile with shape parameters $\tau=0.9$ and $\eta=0.35$.
The shaded area highlights the radial range not accessible to present
numerical simulations.}
\end{figure}

\clearpage
\begin{figure}
\epsscale{1}\plotone{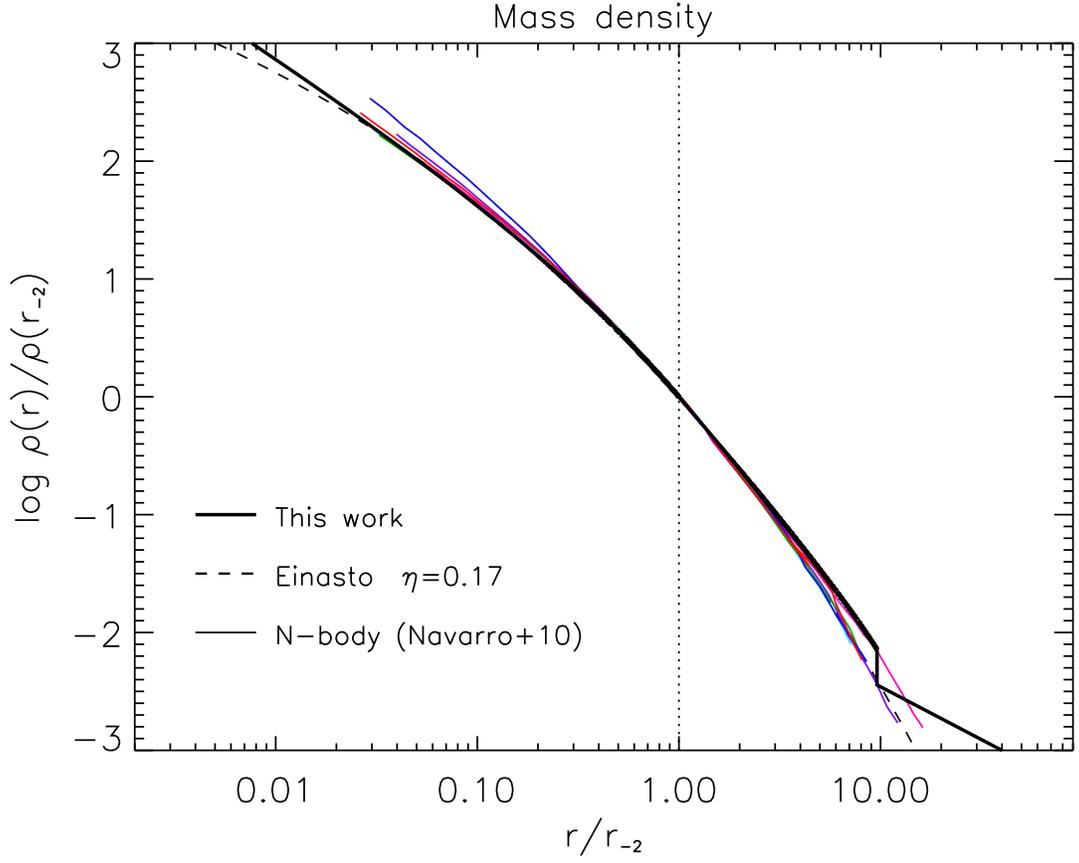}
\caption{Comparison of the self-similar density profiles (thick solid line)
with $\epsilon=1/8$, $\beta_+=0.25$, $\kappa_\star=0.1$ (region inward of
$r_{-2}$) and with $\epsilon=1/2$, $\beta_+=0.25$, $\kappa_\star=0.01$ (region
outward of $r_{-2}$) to the outcomes for six different halos extracted from
the \textsl{Aquarius} $N-$body simulation (Navarro et al. 2010; thin colored
lines); the standard Einasto profile with $\eta=0.17$ is also illustrated for
reference (dashed line).}
\end{figure}

\clearpage
\begin{figure}
\epsscale{1}\plotone{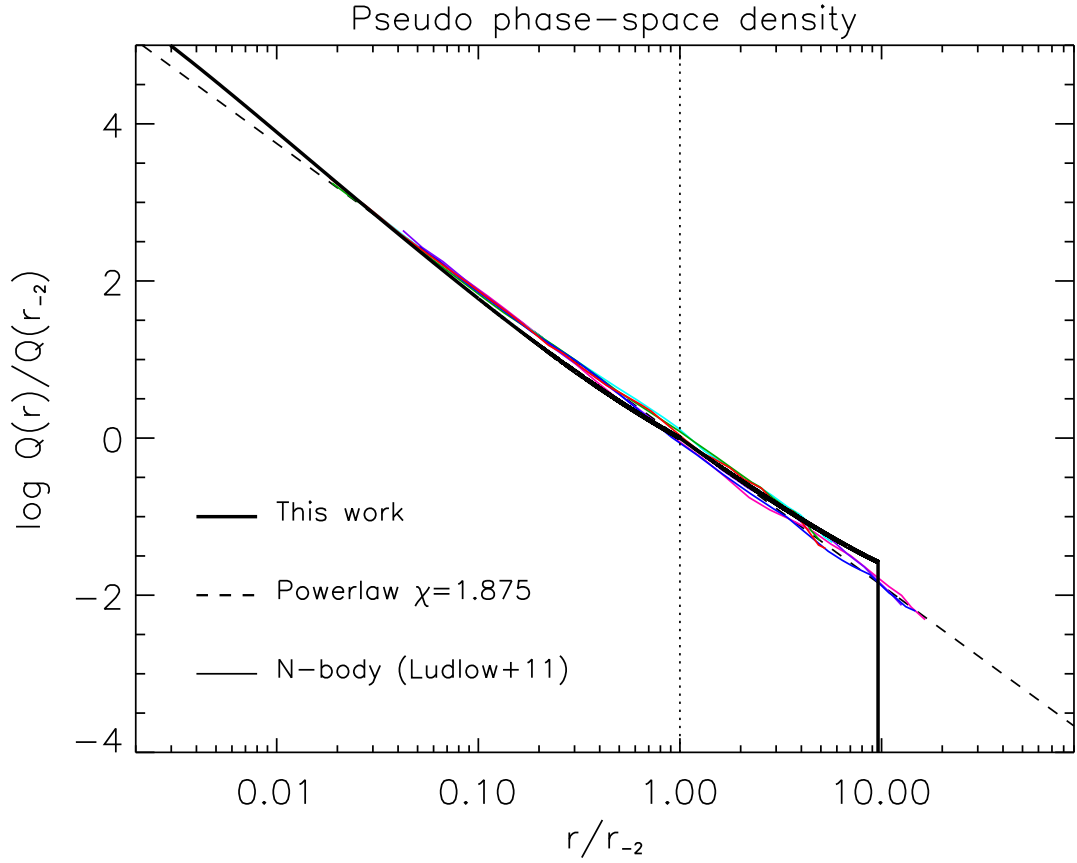}
\caption{Comparison of the self-similar profiles of pseudo phase-space density
(thick solid line) with $\epsilon=1/8$, $\beta_+=0.25$, $\kappa_\star=0.1$
(region inward of $r_{-2}$) and with $\epsilon=1/2$, $\beta_+=0.25$,
$\kappa_\star=0.01$ (region outward of $r_{-2}$) to the outcomes for six
different halos extracted from the \textsl{Aquarius} $N-$body simulation
(Navarro et al. 2010; thin colored lines); the powerlaw $Q\propto r^{-\chi}$
with $\chi=1.875$ is also illustrated for reference (dashed line).}
\end{figure}

\clearpage
\begin{figure}
\epsscale{1}\plotone{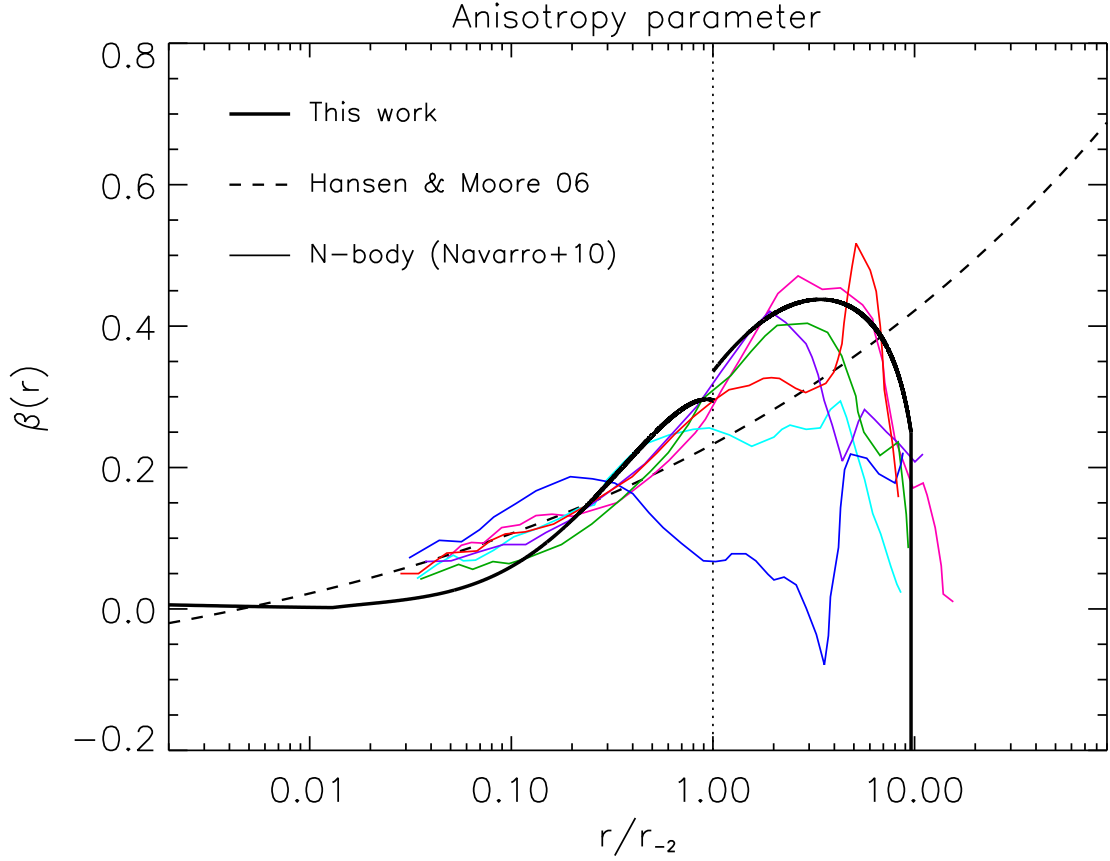} \caption{Comparison of the
self-similar anisotropy profiles (thick solid line) with $\epsilon=1/8$,
$\beta_+=0.25$, $\kappa_\star=0.1$ (region inward of $r_{-2}$) and
with $\epsilon=1/2$, $\beta_+=0.25$, $\kappa_\star=0.01$ (region outward of
$r_{-2}$) to the outcomes for six different halos extracted from the
\textsl{Aquarius} $N-$body simulation (Navarro et al. 2010; thin colored
lines); the anisotropy profile obtained on combining the Einasto density
profile with $\eta=0.17$ and the Hansen \& Moore (2006) $\beta-\gamma$
relation is also illustrated for reference (dashed line).}
\end{figure}

\clearpage
\begin{figure}
\epsscale{1}\plotone{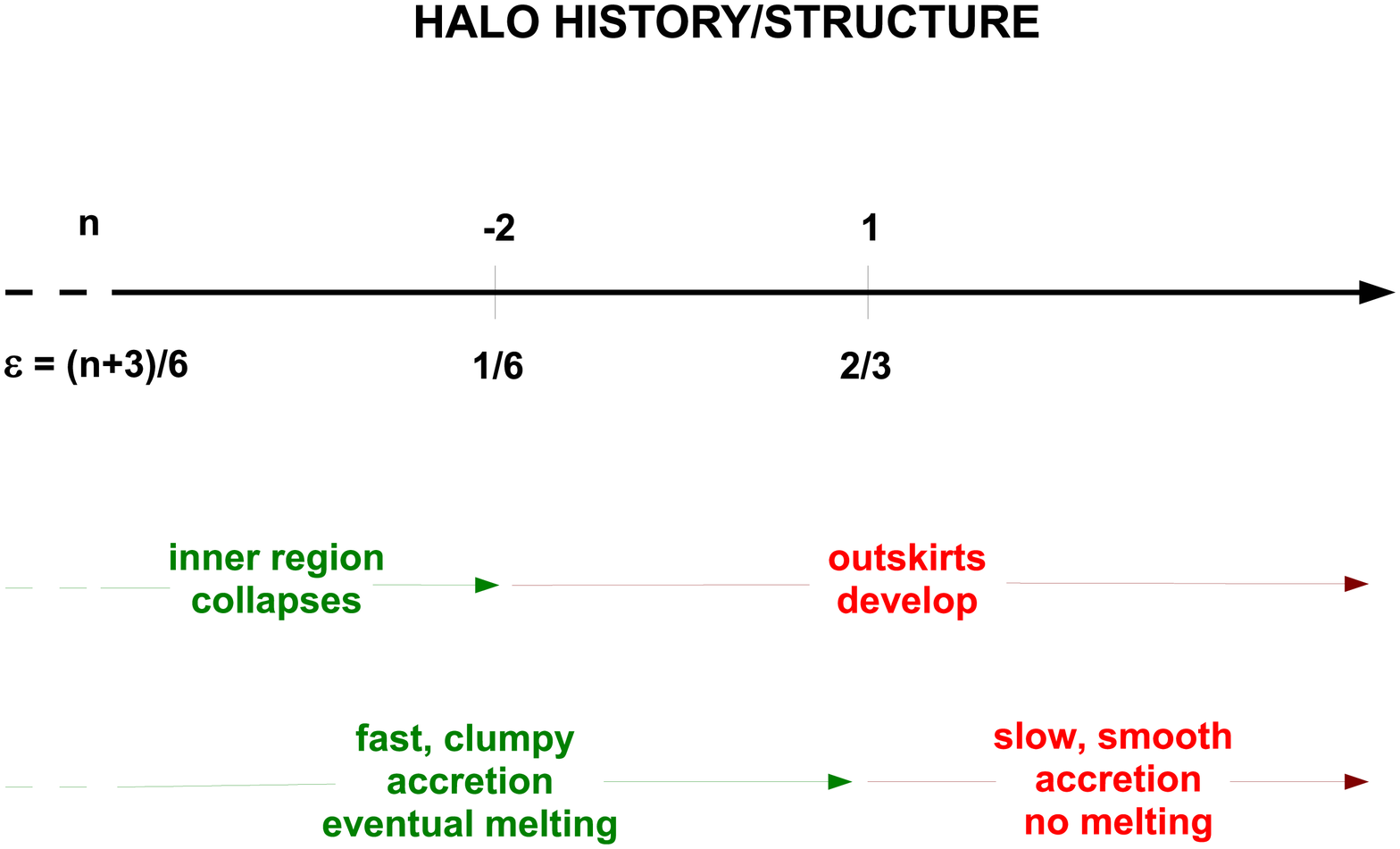} \caption{Schematics of the link
between the halo structure and its two-stage growth history as it emerges
from our self-similar solutions and from numerical simulations (see \S~5 for
details); note that cosmic time runs from left to right.}
\end{figure}

\clearpage
\begin{figure}
\epsscale{1}\plotone{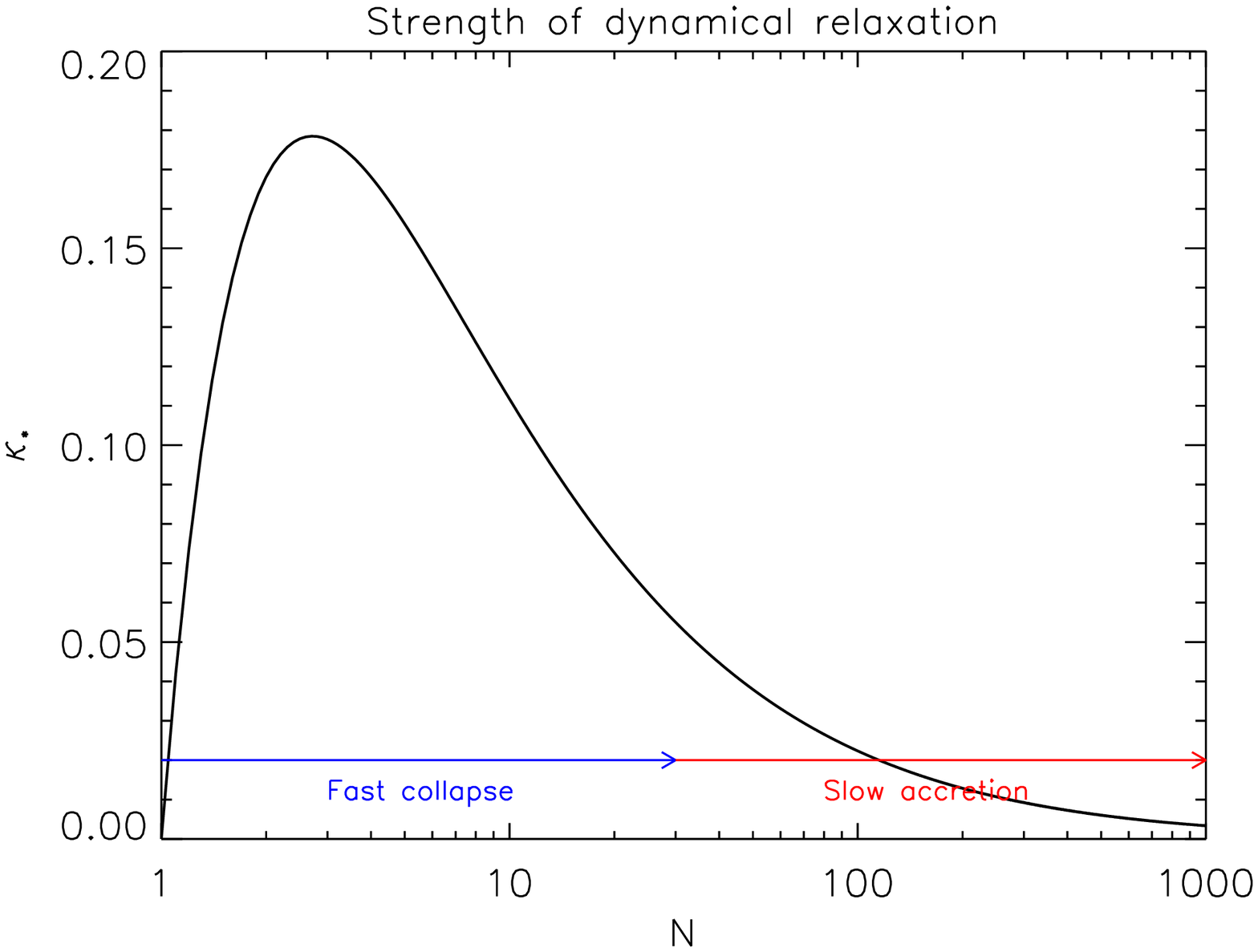}
\figurenum{A1}
\caption{The strength parameter of the dynamical relaxation as a function of
the effective number of clumps $\mathcal{N}\equiv M/m$ in the infalling
matter, see Appendix A for details.}
\end{figure}

\end{document}